\documentclass[citeautoscript,floatfix,aps,prl,onecolumn,superscriptaddress]{revtex4-1}
\usepackage{graphicx}
\usepackage{amsmath}
\usepackage{amssymb}
\usepackage{dcolumn}
\usepackage[version=4]{mhchem}
\usepackage{latexsym}
\usepackage{rotating}
\usepackage{epstopdf}
\usepackage[usenames,dvipsnames]{xcolor}
\usepackage{float}
\usepackage{soul}
\usepackage{epsfig}
\usepackage{psfrag}
\usepackage{natbib}
\usepackage{bm}
\usepackage{eucal}
\usepackage{mathrsfs}
\usepackage{braket}
\usepackage{enumerate}
\usepackage{longtable}
\usepackage{bm}
\usepackage{hyperref}
\usepackage{amsfonts}
\usepackage{dcolumn}% Align table columns on decimal point
\usepackage{bm}
\usepackage{changes}

\newcommand{\be}{\begin{equation}}
	\newcommand{\ee}{\end{equation}}
\newcommand{\bn}{\begin{eqnarray}}
	\newcommand{\en}{\end{eqnarray}}

\def\x2y2{{x^2-y^2}}

\usepackage{color} % use colored text in latex

\usepackage{hyperref}
\hypersetup{
	colorlinks=true,final=true,
	linkcolor=red,
	citecolor=blue,
	filecolor=blue,
	urlcolor=blue,
}
\begin{document}

	\title{A Theory for Colors of Strongly Correlated Electronic Systems}
	\author{Swagata Acharya}
	\affiliation{Institute for Molecules and Materials, Radboud University, {NL-}6525 AJ Nijmegen, The Netherlands}
    \affiliation{National Renewable Energy Laboratory, Golden, CO 80401, USA}
	\email{swagata.acharya@ru.nl}
	\author{Cedric Weber}
	\affiliation{ King's College London, Theory and Simulation of Condensed Matter,
		The Strand, WC2R 2LS London, UK}
	\author{Dimitar Pashov}
	\affiliation{ King's College London, Theory and Simulation of Condensed Matter,
		The Strand, WC2R 2LS London, UK}
	\author{Mark van Schilfgaarde}
	\affiliation{National Renewable Energy Laboratory, Golden, CO 80401, USA}
	\affiliation{ King's College London, Theory and Simulation of Condensed Matter,
		The Strand, WC2R 2LS London, UK}
	\author{Alexander I. Lichtenstein}
	\affiliation{Institute  of  Theoretical  Physics,  University  of  Hamburg,  20355  Hamburg,  Germany}
	\affiliation{European  X-Ray  Free-Electron  Laser  Facility,  Holzkoppel  4,  22869  Schenefeld,  Germany}
	\author{Mikhail I. Katsnelson}
	\affiliation{Institute for Molecules and Materials, Radboud University, {NL-}6525 AJ Nijmegen, The Netherlands}
	%\blfootnote{*Corresponding author:swagata.acharya@ru.nl}

	%\blfootnote{*Corresponding author:swagata.acharya@ru.nl}

	\begin{abstract}

Many strongly correlated transition metal insulators are colored, even though they have large fundamental band gaps and
no quasi-particle excitations in the visible range. Why such insulators possess the colors they do poses a serious
challenge for any many-body theory to reliably pick up the interactions responsible for the color.  We pick two
archetypal cases as examples: NiO with green color and MnF\textsubscript{2} with pink color.  The body of literature
around the collective charge transitions (excitons) that are responsible for the color in these and
  other strongly correlated systems, often fail to disentangle two important factors: what makes them
  form and what makes them optically bright. An adequate answer requires a theoretical approach able to
  compute such excitations in periodic crystals, reliably and without free parameters---a formidable challenge. We
employ two kinds of advanced \emph{ab initio} many body Green's function theories to investigate both optical and spin
susceptibilities.  The first, a perturbative theory based on low-order extensions of the $GW$ approximation, is able to
explain the color in NiO, and indeed well describe the dielectric response over the entire frequency spectrum, while the
same theory is unable to explain why MnF\textsubscript{2} is pink.  We show its color originates from higher order
spin-flip transitions that modify the optical response.  This phenomenon is not captured by low-order perturbation
theory, but it is contained in dynamical mean-field theory (DMFT), which has a dynamical spin-flip vertex that
contributes to the charge susceptibility.  We show that symmetry lowering mechanisms, such as spin-orbit
  coupling, odd-parity phonons and Jan-Teller distortions, determine how `bright' these excitons are, but
  are not fundamental to their existence. As a secondary outcome of this work, we establish that
the one-particle properties of paramagnetic NiO and MnF\textsubscript{2} are both well described by an adequate single
Slater-determinant theory and do not require a dynamical vertex.

\end{abstract}

	\maketitle

	\section*{Introduction}

	\begin{figure}
		\begin{center}
			\includegraphics[width=0.98\columnwidth, angle=-0]{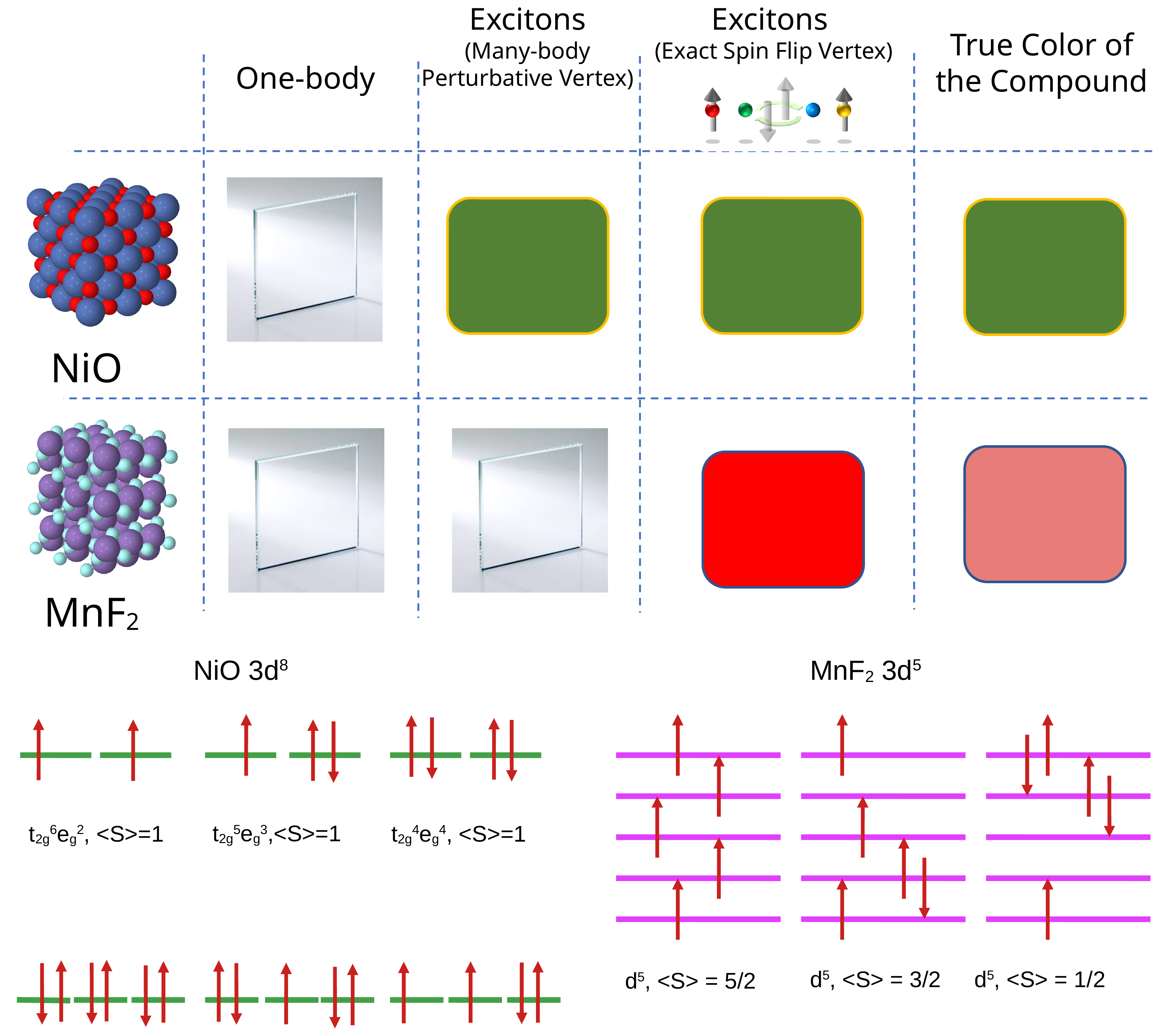}
			\caption{{\bf Fundamentals of the colors in NiO, MnF\textsubscript{2}} : Schematic depiction of fundamental
                          atomic collective charge excitations in NiO and MnF\textsubscript{2}. In NiO $d^8$ to $d^8$ two-particle
                          charge transitions are possible between t$_{2g}$ and e$_{g}$ states that preserve the atomic
                          $\left<S\right>{=}1$ spin configuration. However, in MnF\textsubscript{2} any such collective atomic charge
                          excitations are associated with a simultaneous change in the atomic spin configuration. In
                          NiO, such fundamentally triplet charge excitations determine the green color of the material,
                          while in MnF\textsubscript{2} charge excitations associated with a change in atomic spin configuration
                          determines its pink color. NiO appears green when electron-hole vertex is treated within a
                          paramagnetic formulation of the theory, both nonlocal low-order many-body perturbative and locally
                          exact high-order theory. MnF\textsubscript{2} remains transparent within the paramagnetic many-body perturbative theory,
                          and becomes colored only when the spin-flip component of the electron-hole vertex is
                          incorporated.}
			%The colors are produced for these materials by feeding the respective optical absorption spectra to the java applet color simulator available with the TDDFPT branch~\cite{tddft2} of Quantum Espresso package~\cite{QE-2009,QE-2017,QE-2020}.}
		\label{fig:green}
	\end{center}
\end{figure}

Understanding how light interacts with matter poses an extraordinary challenge. The color of a compound is complementary
to the wavelengths it absorbs; thus an adequate theory of the frequency-dependent dielectric response should be able to
explain the color of real materials.  Often the dielectric response is adequately described in terms of
  an independent particle (Lindhard) description.  Graphite and diamond are respectively black and transparent, because
  the former has no bandgap and absorbs all wavelengths from the visible spectrum, while the bandgap of diamond lies
  outside this range.  The colors of noble metals Cu, Ag, and Au ($d^{10}s^1$) are determined primarily by the
  wavelengths at which photons are reflected, which is controlled by the $d^{10}s^1{\rightarrow}d^9s^2$ transition.  Cu
  has the shallowest $d$ levels, which gives it the copper color, while they are much deeper in Ag and it appears
  silver.  Au is intermediate between the two.  Information for $d^n{\rightarrow}d^{n \pm 1}$ transitions are encoded in
  one-particle Green's functions, and in such cases the absorption should be well characterized, provided the
  one-particle Green's function is adequate and the electron-hole attraction is not strong.  Colors of natural
pigments, found in flowers and vegetables, have been explained~\cite{baroni1,baroni2,baroni3} using time-dependent
density functional theory~\cite{tddft,tddft1,tddft2} (TDDFT) and many body perturbative
approaches~\cite{hedin65,hybertsen85}.  For relatively weakly correlated systems and
molecules~\cite{molecular-spectra,molecular-spectra1,milne2015} such approaches are sufficient to well reproduce the optical
spectrum.  Further, in some rare-earth based fluorosulphide pigments~\cite{jan,galler} the colors have
  been explained recently within a dynamical mean field theory (DMFT) framework.  As for the other systems just
  mentioned, DMFT improves one-particle Green's function, adding higher order local spin fluctuation diagrams missing in
  many body perturbative GW based approaches. Such corrections are important in Hund's metals, Kondo-systems,
  \emph{f}-electron systems, weakly doped Mott insulators.

	\begin{figure}
	\begin{center}
\includegraphics[width=1.0\columnwidth, angle=-0]{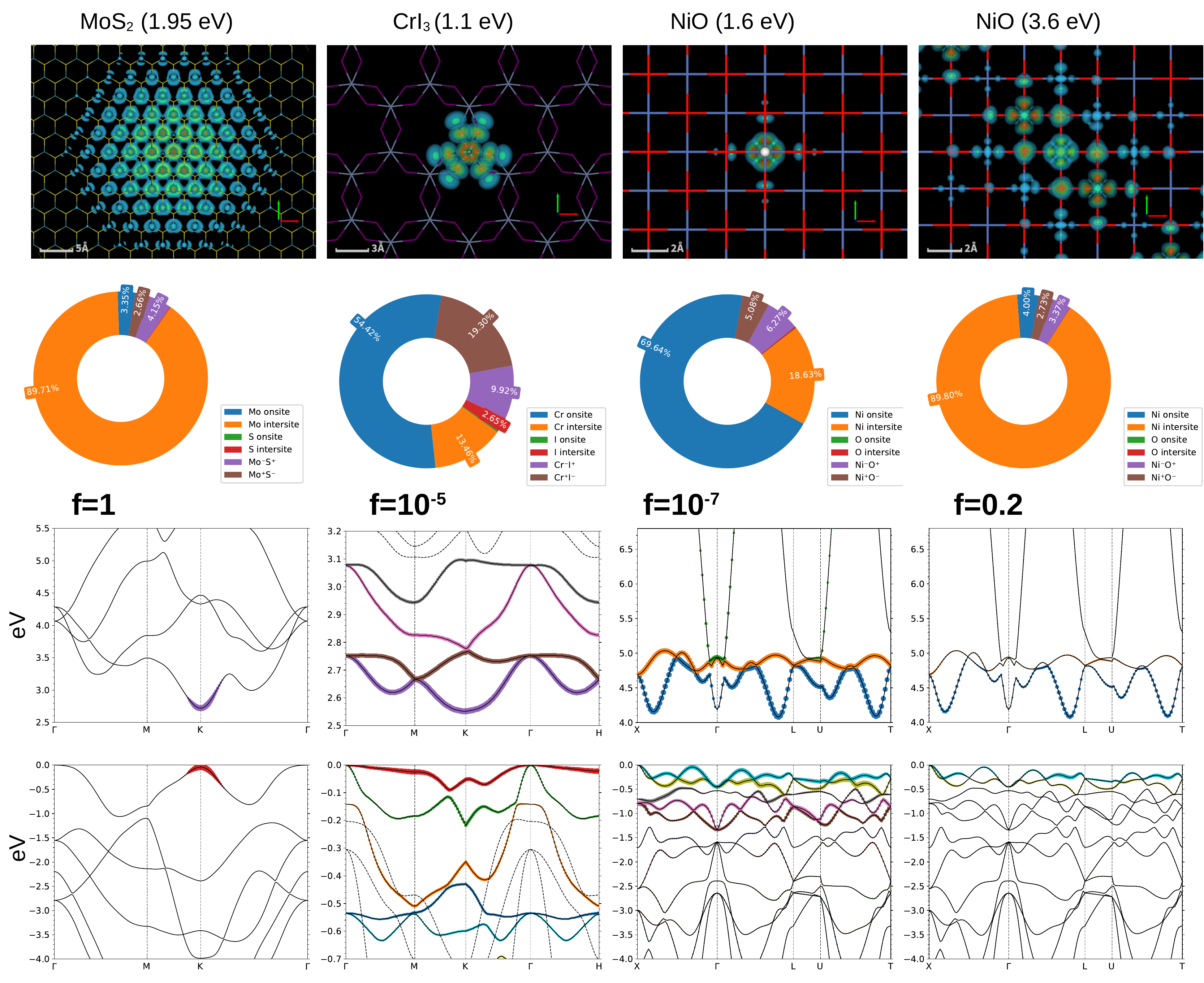}

        \caption{{\bf Orbital, atomic, momentum and real-space decomposition of the excitons:} The
            deepest lying excitons are shown for non-magnetic MoS\textsubscript{2}, ferromagnetic CrI\textsubscript{3} and AFM-NiO.  The
            exciton at 1.95 eV (binding energy E$_{b}$=0.55 eV) in MoS\textsubscript{2} extends to several nanometres while the
            excitons at 1.6 eV (E$_{b}$=2.4 eV) in NiO and at 1.1 eV (E$_{b}$=1.4 eV) in CrI\textsubscript{3}, are localized to
            $\sim$4\,\AA\, only. Bands that are colored take part in the exciton formation, while the width of the bands
            signify relative contributions to the excitonic eigenfunctions. We use different colors to identify
            different bands that take part in the exciton formation.  Several valence and conduction bands and all
            electron and hole momenta contribute almost uniformly to the formation of the Frenkel excitons (at 1.1 eV in
            CrI\textsubscript{3} and 1.6 eV in NiO) while the weakly bound Wannier-Mott excitons (at 1.95 eV in MoS\textsubscript{2} and at
            3.6 eV in NiO) are mostly formed from holes at valence band top and electrons at conduction band bottom. The
            Frenkel excitons are dominantly on-site \emph{d-d} in nature, with sub-leading inter-site \emph{d-d}
            contributions. However, the exciton at the optical shoulder is almost entirely inter-site \emph{d-d} in
            nature. The oscillator strengths ($f$) of the Wannier-Mott excitons are at least four-five orders of
            magnitude higher than the Frenkel excitons. The 1.6 eV exciton in NiO is darker compared to the exciton at
            1.1 eV in CrI\textsubscript{3} because it has more on-site \emph{d-d} component and less \emph{p-d}
            component.}  \label{fig:new} \end{center}
\end{figure}

However, it is well known that the one-particle Green's function may not be enough to
  describe dielectric response.  The vertex in the polarizability is a
    two-particle object responsible for electron-hole attraction which gives rise, e.g. to excitons and optical
  absorption below the fundamental gap~\cite{lucia,louie1,louie2,louie3,Cunningham18,sa-vo2,sa-nbo2}.
When the excitons are weakly bound (Wannier type) they involve mostly states near the edge of the
    fundamental gap: they are confined to a small volume of $k$ space and thus are spread over many lattice sites.
  This is the usual situation in most transition metal dichalcogenides~\cite{louie-bse,louie-bse1,reining-bse}.
The opposite (atomic or Frenkel) limit occurs when the collective excitation takes place between
    highly localized states, typically transitions between different configurations of a particular $\ell$ on an atom,
    e.g. a $d^n{\rightarrow}d^n$ transition.  These atomic-like excitons are governed mostly by properties of the atom
    and depend only weakly on the environment.  For this reason, they have been traditionally tackled via ligand field
    theory: an atom with a few ligands are attached to approximate the solid, and a relatively high-level
    quantum-chemical calculation performed.  Atomic transitions that violate the $\Delta\ell{=}{\pm}1$ selection rule
    are dipole forbidden (Laporte rule), and require some symmetry-breaking mechanism to occur, such as \emph{p-d}
  hybridization, spin-orbit coupling, phonons and Jan-Teller distortions.  Ligand-field theory has been
    a successful traditional line of attack, but it suffers from two difficulties.  First, the brightness of the
    transition depends on the details of symmetry-breaking mechanism, which can be different in the solid than in the
    molecule, and thus affects the brightness of the exciton.  CrI\textsubscript{3} is a prime example: the 2D and bulk
    compounds have similar exciton energies but very different brightness.  Moreover, ligand field theory cannot
    adequately capture excitons larger than the cluster size, and the artificial confinement affects the energy.  This
    occurs when excitons are not entirely Frenkel-like.  (Below we present an exciton in NiO as one instance of this.)
  In such situations, it is crucial to build a parameter-free \emph{ab-initio} theory able to reliably compute both one-
  and two-particle properties in the thermodynamic limit, to give a systematic understanding of what originates these
  excitons and what makes them bright.

CrX\textsubscript{3} is one system where excitons have recently been well characterized
    by low-order many-body perturbation theory~\cite{cri3,acharya2022,crbr3}.  Adding ladder diagrams to the RPA
    polarizability via a Bethe-Salpeter equation (BSE) in the particle-hole charge channel, yields the
  deep Frenkel excitons observed.  To accomplish this, the one-particle Green's function must be also of
    high fidelity.  In Refs.~\cite{cri3,crbr3} the LDA Green's function was augmented by a semiempirical Hubbard
    parameter, which yields improved energy levels but a somewhat inadequate description of the eigenfunctions; in
    Ref.~\cite{acharya2022} a fully \emph{ab initio} approach was employed.  The latter technique, which we also use in
    the present work, extends the $GW$ approximation to include ladder diagrams in \emph{W} for the self-energy.
    Excitonic levels are similar in the two approaches, but the better eigenfunctions in the \emph{ab initio} approach
    give rise to some differences, as will be described elsewhere~\cite{novoselovcri3}.

CrX\textsubscript{3} is fully spin-polarized (t$_{2g\uparrow}^{3}$ with 3 $\mu_{B}$/Cr atom)
  ferromagnetic insulator. The valence and conduction band edges are 3\emph{d}-t$_{2g\uparrow}$ and e$_{g\uparrow}$
  respectively which can host trivially triplet electron-hole excitations between t$_{2g}$ and e$_{g}$ states still
  keeping the atomic $d^{3}$ configuration. These heavily bound Frenkel excitons originate from a
  $d^{3}{\rightarrow}d^{3}$ transition and should be dark according to the Laporte rule, but we showed
  that they became bright in periodic crystals, primarily because the valence and conduction 3\emph{d}
  states strongly hybridize with X-\emph{p} states~\cite{acharya2022}.  Spin-orbit coupling~\cite{suganobook},
  odd-parity phonons, Jan-Teller distortions~\cite{seyler}, and as we show here, spin disorder, can also lower the
  symmetries of the exciton wavefunctions can make them still more bright.

All the excitons in CrX\textsubscript{3} can be picked up by a \emph{GW}+BSE
  approach since spin fluctuations are not
    involved. However, that need not be the case for antiferromagnetic insulators.  The insulating
  band gap of NiO is $\sim$4 eV~\cite{sawatzky-nio} However, NiO is pale green, which implies that at
    least a portion of the light-matter interaction originates from excitons that absorb selectively in the visible part
    of the optical spectrum (380-700 nm or 3.26 eV to 1.65 eV).
  Furthermore, NiO appears green in its bulk crystalline variant, in thin-film and in powdered form, suggesting the
  presence of the a deep lying excitonic absorption common to all forms.  An even more
  dramatic situation emerges in Mott insulating MnF\textsubscript{2}, where the one-particle band gap is $\sim$8
  eV~\cite{mnf2-gap1,mnf2-gap2,mnf2-exciton1}, nevertheless, the material appears pink in its bulk crystalline
  variant. In NiO, Ni assumes a $d^{8}$ atomic configuration (t$_{2g}^{6}d_{z^2}^{1}d_{x^2-y^2}^{1}$) in the solid,
  while Mn is $d^{5}$ in MnF\textsubscript{2}, all electrons with the same spin (Fig.~\ref{fig:green}). In analogy with CrX\textsubscript{3},
  in NiO, the Frenkel excitons with atomic $d^{8}{\rightarrow}d^{8}$ transitions should still be picked up in many-body
  perturbative framework since an essentially atomic triplet exciton can emerge without requiring a
    spin flip. This is accomplished by creating a hole in the t$_{2g}$ orbitals and adding an electron in the
  half-filled e$_{g}$ orbitals, which can preserve the $\left<S\right>{=}1$ ground state atomic configuration of Nickel
  (see Fig.~\ref{fig:green}).  A prior theoretical work \cite{claudia} did study excitons above 3\,eV
    but such excitons do not explain the color in NiO.  In fully antiferromagnetically ordered NiO such triplet transitions should be
  dark according to the Laporte rule; however, as we have already discussed, they
    nevertheless exist, and moreover can become bright owing to a symmetry-lowering mechanism, as we
    will discuss below.  The situation is completely different in MnF\textsubscript{2}, where any atomic
  electron-hole transition, $d^{5}{\rightarrow}d^{5}$, necessarily involves a spin-flip mechanism (see
  Fig.~\ref{fig:green}). Such excitons are completely absent from a standard \emph{GW}+BSE framework and needs a higher
  level theory which incorporates local spin fluctuations that modify the charge component of the
  two-particle Green's function.

We will employ two approaches: a the \emph{GW}+BSE perturbative approach and an
  augmentation of \emph{GW} with dynamical mean field theory (DMFT).  A brief description of both can be
    found in the Methods section below; see also Refs.~\cite{Kotani07,questaal_paper,cunningham21} for a description of MBPT,
    and Refs.~\cite{questaal_paper,acharya2019} for DMFT.
    MBPT includes non-local charge correlations but misses out on spin fluctuation diagrams,
while in DMFT nonlocality is limited to orbitals on an atomic site.  Within that
    restriction it is locally exact and incorporates all local spin fluctuation
  diagrams. We will establish that the one-particle properties of both NiO and MnF\textsubscript{2}
  are equally well described by either approach. However, the perturbative approach fails
 to explain excitons in MnF\textsubscript{2} and needs the locally exact method with spin fluctuation
  diagrams to yield its pink color. In the process, we will discuss the crucial technical ingredients
  of such theories and its ability in predicting collective responses in strongly correlated systems. We will also show
  that our theory can compute reliably the eigenvalues and eigenfunctions of all such fundamentally dark excitons in
  anti-ferromagnetic insulators, however, symmetry-lowering-mechanisms that are present in a real materials are crucial
  in determining the `brightness' of these excitons.  Using DMFT, we will explore the scaling behavior of the
  eigenvalues of these excitons with Hund's coupling. We will establish that the hardest problem is to explain the
  interactions that are responsible for the pink color in MnF\textsubscript{2}, since it is a strictly half-filled
  antiferromagnet. However, we are able to solve the problem in its entirety and will establish that spin-flip vertex
  that knows about the physical spin-disordering mechanism of the atomic multiplet structure is responsible for its
  color.

\section*{Results and Discussion}

\subsection*{Many-body perturbative theory: `dark' excitons in the visible range in antiferromagnetic NiO:}

We begin with a study of NiO, and employ a newly developed many-body perturbative
Green's function \emph{GW} theory QS$G\widehat{W}$~\cite{Cunningham18,cunningham21,swagcrx1,acharya2022}, an extension of the quasiparticle self-consistent \emph{GW} approximation
(QS\emph{GW})~\cite{mark06qsgw,Kotani07,questaal_paper}, where the polarizability needed to construct \emph{W} is
computed including vertex corrections (ladder diagrams) by solving a Bethe-Salpeter equation (BSE) for the two-particle
Hamiltonian~\cite{cunningham21}.  There are some crucial differences in the implementation of QS$G\widehat{W}$ presented
here from other widely used implementations of BSE. Self-consistency in both $\Sigma$ and the charge
  density~\cite{swagcrx1,swagfirst} plays a crucial role: the theory becomes uniformly consistent so that uncontrolled
  \emph{ad hoc} empirical additions are not needed.  Computing the screened coulomb interaction $\widehat{W}(\omega,q)$
 with ladders included in the polarizability largely eliminates the tendency for
  QS\emph{GW} to overestimate the fundamental bandgap. (The vertex in the construction of $\widehat{W}$ is approximated
with a static RPA $W$, as is customary~\cite{louie-bse,louie-bse1,reining-bse}).  \emph{G}, $\Sigma$ and $\widehat{W}$
are iterated until all of them converge.  In each cycle, $W$, is made anew, and the four-point polarizability is
computed to make $\widehat{W}$ from an updated $W$.

\begin{figure}
	\begin{center}
		\includegraphics[width=0.98\columnwidth, angle=-0]{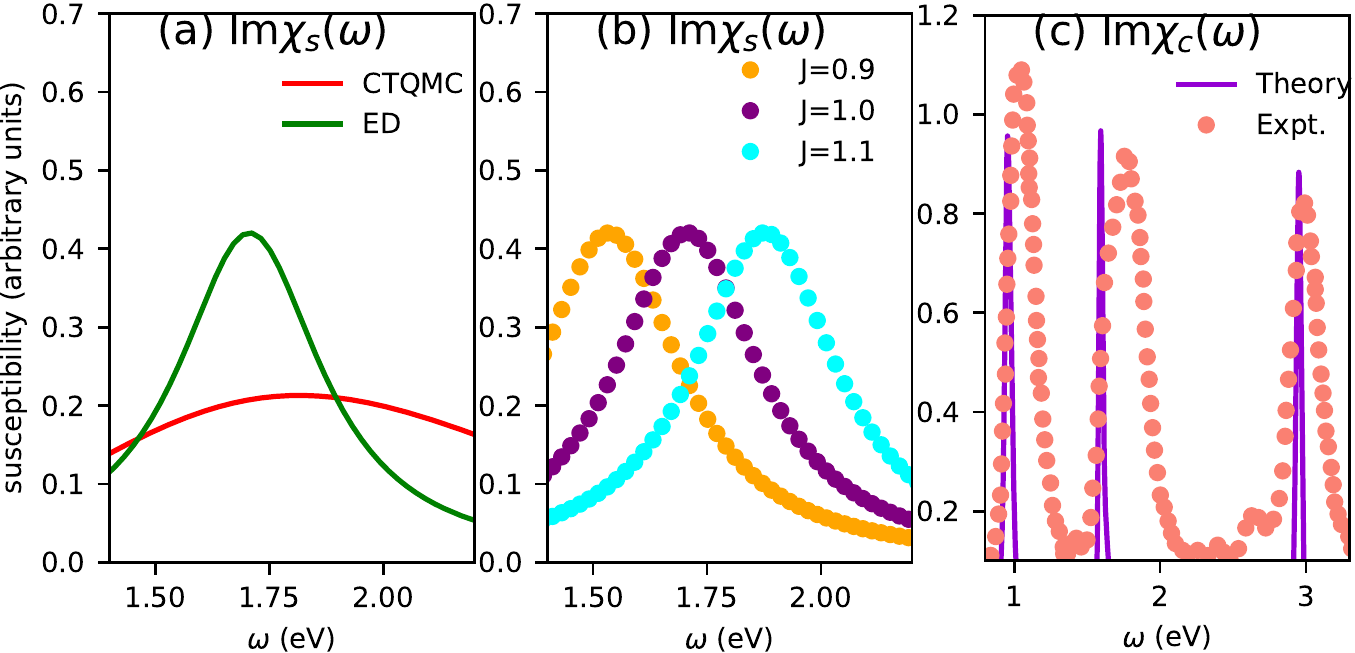}
		\caption{{\bf Peaks in spin and charge susceptibilities at different energies from an exact many-body approach:} : (a) Imaginary part of dynamic spin
			susceptibility Im$\chi_{s}(\omega)$ computed in paramagnetic DMFT, in presence of local vertex
			corrections, show a collective spin excitation at 1.7 eV. The energy of the excitation is robust
			across choices of different impurity solvers (ED or CTQMC) in DMFT. This peak is associated with the
			triplet to singlet transition on the atomic site. (b) The energy of this peak scales with Hund's
			coupling $J$ as 1.7\,$J$, suggesting absence of this spin transition for $J$\,=0. (c)
			Experimentally observed peaks in the collective excitations as measured by
			non-resonant inelastic X-ray scattering measurements~\cite{xray1}, which are insensitive to optical
			matrix elements, are plotted alongside the charge susceptibility computed from DMFT. The intensity of
			the charge peaks in DMFT are multiplied by a large factor to bring them to same scale with
			experiment. The number and the energy of the peaks agree remarkably well with the experiment.}
		\label{fig:spincharge}
	\end{center}
\end{figure}

With a self-consistent self-energy $\Sigma{=}iG\widehat{W}$ in hand, we use the BSE to compute the macroscopic
dielectric response function in NiO, initially modeling the antiferromagnetically (AFM) ordered phase.  Within
QS$G\widehat{W}$, NiO has 4.0\,eV fundamental band gap compared to 5.0\,eV~\cite{supple} from QS\emph{GW}~\cite{nio-4.8,cunningham21}.  This shows that
the vertex significantly modifies the self-energy in NiO~\cite{swagcrx1,Cunningham18}, softening \emph{W} and thus
reducing the gap.  The seminal work by Sawatzky and Allen~\cite{sawatzky-nio} estimated the gap to be $\sim$4.3 eV from
Bremsstrahlung inverse spectroscopy (BIS). Our QS$G\widehat{W}$ one-particle spectra agrees remarkably well with the BIS
data~\cite{cunningham21}.  A recent XPS study caps the gap at 4.0 eV (within 0.35 eV energy
resolution)~\cite{nio-4ev}. Next, we compute the vertex corrected macroscopic dielectric response functions and find
that there are in-gap optical absorption down to 3.6 eV, i.e. $E_{b}{\sim}0.4$\,eV deeper
than the edge of the one-particle band gap in QS$G\widehat{W}$. Our computed imaginary part of macroscopic dielectric
response $\epsilon_{2}$ agrees remarkably well~\cite{cunningham21} with the experimental data from Powell and
Spicer~\cite{spicer-optics} over the full frequency range. This shows with a suitably vertex corrected self-energy,
QS$G\widehat{W}$ incorporates the relevant electron-hole vertex corrections that can red-shift the optical spectral
weight by a significant amount. A previous semi-empirical GW+BSE calculation~\cite{claudia} shows similar optical
absorption inside band gap around 3.6 eV.  Nevertheless, such optical absorption does not produce the
right color~\cite{supple}.  Moreover, without phenomenological adjustments such as was done in
  Ref.~\cite{claudia}, single-shot calculations often have limited success in describing the optical absorption in
  antiferromagnets.  Nor is it sufficient to well describe the one-particle spectrum: the electronic eigenfunctions are
  biased by the density-functional eigenfunctions and yield poor description of their orbital character.
We discuss this aspect in detail in the supplemental materials and show explicitly how our
  self-consistent vertex corrected approach systematically corrects the electronic eigenfunctions in a fully
  diagrammatic fashion without the need of \emph{ad hoc} parameters.

We find strong optical absorption starting at 3.6\,eV, as we have discussed
    elsewhere~\cite{cunningham21}.  We also  find a deep lying eigenvalue of
  the two-particle e-h Hamiltonian at $\sim$1.6 eV, but  the
  oscillator strength of this eigenvalue is five orders of magnitude smaller than the optical shoulder at 3.6 eV (see
  Fig.~\ref{fig:new}).  This exciton has the right energy to absorb the red part of the visible
  spectrum, thereby causing NiO to appear green.  We resolve the corresponding eigenfunction in the atomic, orbital, band basis and also in
  real-space (see Fig.~\ref{fig:new}). Nearly 70\% of the spectral weight comes from on-site and
  $\sim$20\% comes from inter-site e$_{g}$-t$_{2g}$ transitions. The rest of the spectral weight comes from e-h
  processes shared between Ni and O. The situation is dramatically different for absorption from the 3.6\,eV
  exciton. The inter-site e$_{g}$-t$_{2g}$ transition becomes the most dominant mechanism followed by processes where
  electron and holes are shared between Ni and O. The Ni on-site element almost entirely vanishes at optical shoulder,
  and the oscillator strength, as a consequence,  is $\sim$5 orders of magnitude
larger than the 1.6 eV exciton.

To make this exciton bright, some symmetry-breaking mechanism is needed. Spin-orbit coupling (SOC)
(not included in our calculation) is one obvious candidate; indeed  the classic works by Sugano and
  Tanabe~\cite{sugano1954,suganobook} used a combination of atomic ligand field theory and SOC as the main theoretical
  foundation that relaxes atomic Laporte rule to explain excitonic absorption in several transition metal oxide
  insulators. However, other mechanisms can enhance brightness as well. In extended systems
 there is a finite \emph{d-p} hybridization, as we showed recently for two-dimensional magnetic semiconductors CrX\textsubscript{3}~\cite{acharya2022,novoselovcri3}. Contrasting NiO and CrX\textsubscript{3} sheds light on their differences:  the
  deepest lying bright exciton (at 1.1 eV) in ferromagnetic monolayer of CrI\textsubscript{3} has two orders of magnitude higher
  oscillator strength compared to the 1.6 eV exciton in AFM NiO. As Fig.~\ref{fig:new} shows, the excitonic spectral
  weight in CrI\textsubscript{3} has nearly 30\% contribution coming from processes where electron and holes are shared between Cr
  and I, in strong contrast to 1.6 eV exciton in NiO that has only about 10\% contribution from such processes. Both of
  them are fundamentally triplet and
mostly on-site in nature. Computing them both in their crystalline environment without assumptions behind ligand-field theory
    makes it possible to compare the role of the lattice, and moreover show explicitly the quantitative nature of the
  difference in their spectral decomposition. This is a step change in the study of collective charge excitations in
  antiferromagnetic insulators.  Having said that, we will show in the following sections the primary processes that are
  responsible for making these excitons `brighter', without having to introduce \emph{ad hoc} arguments for the symmetry
  lowering mechanism.

\subsection*{Many-body perturbative theory: `bright' excitons in the visible range in paramagnetic NiO:}

Spin disorder can be a mechanism to turn dark excitons bright.  To show this we construct a Special QuasiRandom
Structure~\cite{sqs}, a 2$\times$2$\times$2 supercell of the AFM structure consisting of 16 Ni in a (pseudo)paramagnetic
spin arrangement and compute the self-energy with QS$G\widehat{W}$.  The self-consistent PM and AFM local moments are
very similar ($|M|{=}1.76{\pm}0.01\,\mu_B$ and $1.72\,\mu_B$ in the two cases), and the PM band structure strongly
resembles the AFM when folded to a 2$\times$2$\times$2 supercell.  The QS$G\widehat{W}$ (QS\emph{GW}) gap~\cite{supple}, differs from the AFM gap by 0.2\,eV.  This flatly contradicts the commonly made
assertion~\cite{Brandow75} that NiO cannot be described in terms of a conventional Slater (band) theory, which
``attribute[s] the insulating gap to the existence of long-range magnetic order \cite{Ren06}'', and requires a dynamical
theory such as DMFT for an adequate description of the electronic structure.  The input for a DMFT calculation is
usually a non-magnetic metallic phase, a charge (Mott) gap opens because of dynamically fluctuating local moments,
according to the theory.  The present calculations establish conclusively that local magnetic moments are essential to
form the gap, but that dynamical fluctuations are not. A similar conclusion was reached in a recent work on Mott
insulators~\cite{zunger2}. Thus, the formation of the gap can be equally represented by a frequency-dependent
site-local potential such as used in DMFT, or a low-order perturbative approach like \emph{GW}. The latter has important
advantages: it is fully \emph{ab initio} and simpler, without the complications of partitioning into a correlated
subspace with a parameterized hamiltonian; and it well describes optical excitations over a wide energy
window. The single-most important contribution of spin-disordering is that it makes the 1.6 eV exciton
  bright, enhancing its oscillator strength by at least three orders of magnitude compared to the fully ordered ideal
  AFM situation. The matrix elements between up- and down- spins do not vanish in the paramagnetic state, making this
peak optically bright (see Fig.~\ref{fig:spincharge}).  While the fully disordered limit is applicable
  only above the N\'eel temperature, spin fluctuations occur at all temperatures~\cite{Fisher59}: thus they can be an
  important contributor to the brightness of the 1.6\,eV exciton.  In a future work, we will assess the relative
  importance of this effect and spin orbit coupling on the colour in NiO.

\subsection*{Many body locally exact theory: excitons in the visible range from paramagnetic DMFT in NiO:}

In a parallel approach, we perform a conventional non-magnetic QS\emph{GW}+paramagnetic DMFT~\cite{antoine-rmp}. The
impurity Hamiltonian is built out of the Ni-3\emph{d} orbitals, which hybridize with the bath.  All five Ni-3\emph{d}
orbitals which are comprised of t$_{2g}$ and e$_{g}$ states, are kept in the Hubbard Hamiltonian.  The Anderson impurity
model is solved in the presence of a rotationally invariant interaction matrix. We find that the Mott-Hubbard gap opens
through DMFT self-consistency and the band gap is $\sim$4 eV for U=8 eV and $J$=1.0~\cite{sasha-nio}. We solve the
Anderson impurity model with two different exact impurity solvers, exact-diagonalisation (ED) and continuous-time
Quantum Monte Carlo (CTQMC)~\cite{hauleqmc,gull}, and find that all the essential conclusions are independent of the
choice of the solver. We compute the local spin ($\chi_{s}$) and charge ($\chi_{c}$)
susceptibilities on real frequencies from ED and we resolve them in their intra- and inter-orbital components. In Fig.~\ref{fig:spincharge} (a) we show that the energy of the peak in $\chi_{s}$ is essentially same from CTQMC and ED.  Within
paramagnetic DMFT, the magnetic vertex can be written as $\Gamma_{s}$ =
$\Gamma_{\uparrow\uparrow}$-$\Gamma_{\uparrow\downarrow}$ and the charge vertex as $\Gamma_{c}$ =
$\Gamma_{\uparrow\uparrow}$+$\Gamma_{\uparrow\downarrow}$. It is only natural that spin and charge susceptibility peaks
appear at different energies in paramagnetic DMFT. DMFT local irreducible vertex functions are site-local, in contrast
to the non-local static vertex in QS$G\widehat{W}$. The most remarkable difference between the two is the spin-flip
component of the vertex in DMFT, $\Gamma_{\uparrow\downarrow}$, which is completely absent within QS$G\widehat{W}$.

\begin{figure}
	\begin{center}
		\includegraphics[width=0.98\columnwidth, angle=-0]{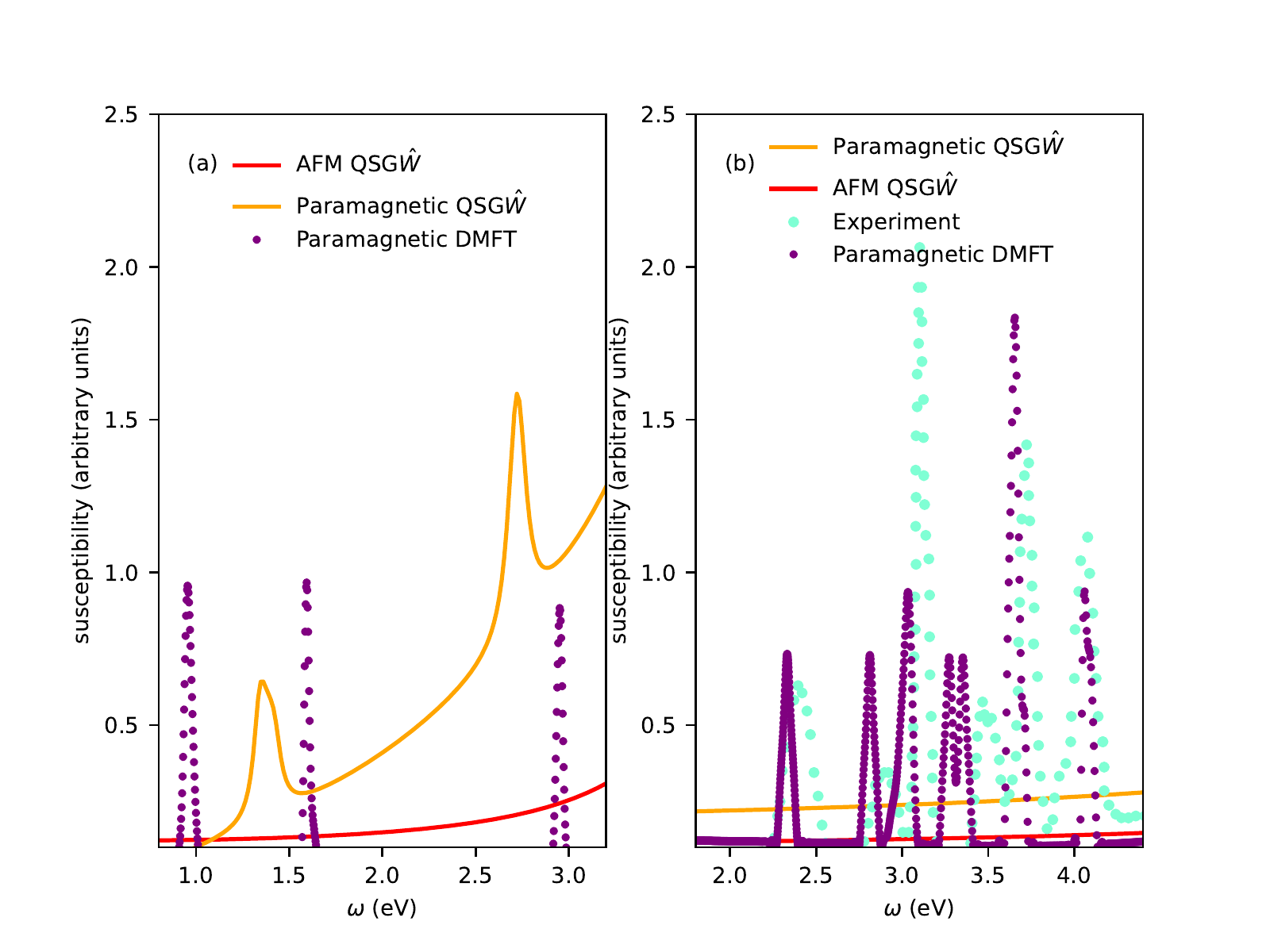}
		\caption{{\bf Colors in NiO and MnF\textsubscript{2} determined by the essential structure of the electron-hole vertex from perturbative and exact many-body approaches}: (a) We show the imaginary part of the
			charge susceptibility $\chi_{c}$ computed from different levels of the theory and from experiment. In
			NiO, both paramagnetic QS$G\widehat{W}$ and paramagnetic DMFT can produce the right color of the material as
			they account for the peaks at $\sim$1.6 eV and $\sim$3 eV that essentially absorbs the red and blue
			part of the visible spectrum respectively and can make NiO appear green. The missing peak from
			paramagnetic QS$G\widehat{W}$, compared to DMFT, at $\sim$1 eV is outside the visible range and does not
			contribute to the color of the material. (b) In MnF\textsubscript{2} it is the spin flip
			component of the DMFT charge vertex that is essential to the collective electron-hole charge
			excitation on the atomic site of Mn. So the collective excitations are only observed in DMFT and are
			completely absent from both AFM QS$G\widehat{W}$ and paramagnetic QS$G\widehat{W}$. The charge susceptibility peaks as
			observed in DMFT agree remarkably well with the series of optical absorption peaks observed in
			experiments~\cite{stout1959,suganobook}.}
		\label{fig:opticsboth}
	\end{center}
\end{figure}

Three distinct peaks are observed in $\mathrm{Im}\,\chi_{c}$ in the energy window
0--3\,eV: at 0.93\,eV, 1.55\,eV and 2.91\,eV (see Fig.~\ref{fig:spincharge}(c)). All these two-particle charge
excitations preserve the $\left<S\right>{=}1$ atomic configuration on average (see
Fig.~\ref{fig:green}).
%Also we observe that these exciton peaks are only in the inter-orbital components of the $\chi_{c}$ that allows an
%excitation from a t$_{2g}$ orbital to a e$_{g}$ orbital.
The three peaks correspond to t$_{2g}${$\rightarrow$}e$_{g}$ transitions: excitations of the ground
state t$_{2g}^{6}$e$_{g}^{2}{\rightarrow}\mathrm{t}_{2g}^{5}$e$_{g}^{3}$,
t$_{2g}^{6}$e$_{g}^{2}{\rightarrow}\mathrm{t}_{2g}^{4}$e$_{g}^{4}$ and a third which
is combination of the first two. The peaks found when
ED is used to solve the Anderson impurity problem are also found when the CTQMC solver is used; but as the CTQMC bath
is not discretised the peaks are broadened and harder to detect.  Nevertheless, the three-peak structure is robust
and is independent of the DMFT solver.  Intriguingly enough, two of these three peak positions match almost exactly with
the excitons observed in paramagnetic QS$G\widehat{W}$ (see Fig.~\ref{fig:opticsboth}).

\subsection*{Distinct scaling features for peaks in spin and charge susceptibilities with Hund's coupling:}

The spin susceptibility $\chi_s$ is somewhat different.  Irrespective of whether only the (half-filled) e$_{g}$ states
or t$_{2g}$ and e$_{g}$ states are all included in the impurity Hamiltonian, $\mathrm{Im}\,\chi_{s}$ always shows a
single high intensity absorption peak at 1.7\,eV (see Fig.~\ref{fig:spincharge} (a)). In the spin channel, this peak is
purely associated with a triplet-singlet ($\left<S\right>{=}1{\rightarrow}\left<S\right>{=}0$) transition which changes
the ground state from t$_{2g}^{6}$d$_{z^2}^{1}$d$_{x^2-y^2}^{1}$ to t$_{2g}^{6}$d$_{z^2}^{0}$d$_{x^2-y^2}^{2}$ or
t$_{2g}^{6}$d$_{z^2}^{2}$d$_{x^2-y^2}^{0}$, as both of them are equally probable due to the cubic crystal
field). The 1.7\,eV peak scales with Hund's coupling $J$ in a manner completely different from the peak
  in $\mathrm{Im}\,\chi_{c}$ at a similar energy (see Fig.~\ref{fig:spincharge}(b) and Table~\ref{tab:binding})
  suggesting that a coupling between them mediated by SOC that would make this exciton bright would be sufficient but
  not fundamentally necessary. In a pure atomic multiplet transition ($\left<S\right>$=1 to $\left<S\right>$=0) the peak
  should be at $2J$, but it is at $1.7J$ and scales as $1.7J$, as determined by doing calculations varying $J$ as a
  parameter; see Fig.~\ref{fig:spincharge} (b)). This suggests that a purely atomic scenario to explain the fundamental
  optical and magnetic transitions is only partially valid, that there are additional spin and charge fluctuations that
  are correctly incorporated within a paramagnetic calculation. The optical spectrum computed from the paramagnetic
phase either in QS$G\widehat{W}$ and QS\emph{GW}+DMFT, produce the long-sought green color of NiO. The low energy peak
at $\sim$1.6\,eV partially absorbs the red part of the visible spectrum, while the absorption above $\sim$2.8\,eV (see
Fig.~\ref{fig:opticsboth} (a)) partially absorbs the blue part of the visible spectrum, leading to the green color of
NiO (see Fig.~\ref{fig:green}). Note that the peaks in $\mathrm{Im}\,\chi_{c}$ from DMFT are low intensity as they are
dipole-forbidden. Such excitations are entirely absent in a purely one-particle picture.  One-particle Green's functions
allow excitations of the atomic states of the form 3\emph{d}$^{8}{\rightarrow}3\mathrm{d}^{7}$ or
3\emph{d}$^{8}{\rightarrow}3\mathrm{d}^{9}$ (one electron on the atomic site goes into the bath or arrives from the
bath).

These low-energy peaks emerge from the local vertex in the two-particle Green's function that allows
3\emph{d}$^{8}{\rightarrow}3d^{8}$ transitions.  These peaks can become bright if any symmetry breaking mechanism such
as spin-orbit coupling is present.  However, as we discuss above, SOC is not fundamental to these transitions, nor to
the essential charge absorption that determines the color. Our observations are fully supported by
  recent X-ray measurements~\cite{xray2,haverkort,xray1}, which is insensitive to the optical matrix elements, where all
  these three peaks are observed at almost the same energies as our theory (see Fig.~\ref{fig:spincharge}
  (c)). Intriguingly enough, in the classic experimental work by Propach and Reinen~\cite{propach1978}, they picked up
  all the three peaks in the optical emission spectra.  Powell and Spicer~\cite{spicer-optics}, however, failed to see
  these three peaks in their optical absorption spectra. This is only natural since e-h emission processes can couple to
  non-radiative mechanisms present in the crystal and that can make the emission lines bright enough to be picked up in
  spectroscopy. In essence, this re-affirms our theoretical results that these excitons primarily emerge from the
  essential structure of the two-particle electronic interactions (vertex) and it is their brightness that is determined
  by spin-orbit coupling or coupling of these excitons to different bosons.

Further, linear scaling of the peak in Im$\chi_{s}$ also implies that as $J{\rightarrow}0$, the peak in spin
susceptibility must vanish. However, no such scaling holds for the charge peak at $\sim$1.5\,eV: its position changes
little as $J$ is varied. This is another signature for the charge peaks, that distinguishes them from the spin.  We find
the 0.93\,eV and 1.55\,eV peaks to be independent of $J$ and the 2.9\,eV peak to scale with $J$ as
1.40\,eV\,+\,1.5\,$J$. All three charge absorption peaks should survive even when $J$ is 0.  The $\sim$1\,eV peak seen
in DMFT but absent from QS$G\widehat{W}$, originates primarily from to the dynamics in the vertex in DMFT, which was
omitted from the QS$G\widehat{W}$ vertex. However, this low-energy peak does not contribute to the color of NiO as it is
far outside the visible range.

\begin{table}[h]
	%  \begin{center}
		\footnotesize
		\begin{tabular}{c|@{\hskip 3pt}ccc@{\hskip 3pt}|@{\hskip 3pt}c}
			\hline
			theory & \multicolumn{3}{c}{charge peaks (eV)} & \multicolumn{1}{c}{spin peaks (eV)} \cr
			%			& \ce{peak 1} & \ce{peak 2}  & \ce{peak 3} & \ce{peak 3}  \cr
			\hline
			$J$=0.8\,eV   & 0.93   & 1.52   & 2.6   & 1.36  \cr
			$J$=0.9\,eV  & 0.93  & 1.535  & 2.76   &  1.53   \cr
			$J$=1.0\,eV   & 0.93  & 1.55  & 2.91   & 1.7  \cr
			$J$=1.1\,eV   & 0.93  & 1.57  & 3.06   & 1.87  \cr
		\end{tabular}
		%  \end{center}
	\caption{{\bf Distinct scaling features of the peak energies in charge and spin susceptibility Im$\chi_{c}$ and
            Im$\chi_{s}$ with Hund's coupling $J$}.  The peak labelled ``spin peak'' is computed from Im$\chi_{s}$ and
          scales as $1.7J$.  Peaks in Im$\chi_{c}$ (labelled ``charge peak'') scale differently with $J$. In particular,
          the exciton peak at 1.6 eV (essential for the green color in NiO), is insensitive to $J$, suggesting no cross
          coupling with $\chi_{s}$, though a peak is present in both $\chi_{s}$ and $\chi_{c}$ at a similar wavelength.
	}
	\label{tab:binding}
\end{table}

\subsection*{Many-body perturbative theory: absence of excitons in the visible range in both antiferromagnetic and paramagnetic MnF\textsubscript{2}:}

We now turn to MnF\textsubscript{2}. The reduced symmetry modifies the crystal field: now the five Mn \emph{d} states
are non-degenerate and each is half-filled (see Fig.~\ref{fig:green}). This differs qualitatively from NiO, with its
filled degenerate t$_{2g}$ and half-filled degenerate e$_{g}$ levels. The scenario for possible optical transitions is
thus much wider in MnF\textsubscript{2}, while on the other hand, any $d^{5}{\rightarrow}d^{5}$ transition requires a spin flip.
When AFM QS\emph{GW} is compared to paramagnetic QS\emph{GW}, the one particle gaps are slightly different
(9.1\,eV for the AFM case and 8.6\,eV for the PM case, QS$G\widehat{W}$ gap for AFM case is 8.4\,eV), similar to NiO~\cite{supple}. These numbers are in the right
ballpark with available experiments (the fundamental gap for MnF\textsubscript{2} is not very well known and
experiments~\cite{mnf21,mnf22,mnf23} report numbers in the range of 8 to 10\,eV).  For the charge susceptibility the
lowest eigenvalue of the two-particle e-h Hamiltonian found to occur at 6.2\,eV, for both AFM and PM cases.  For the
present purposes, the essential point is that the lowest eigenvalue of the QS$G\widehat{W}$ two-particle hamiltonian is
outside the visible (6.2\,eV) which predicts MnF\textsubscript{2} to be colorless, inconsistent with the faded pink
color observed.
%AFM QS$G\widehat{W}$ and paramagnetic QS$G\widehat{W}$, we observe that the lowest eigenvalue of the two-particle e-h
%Hamiltonian is at the same energy, 6.2\,eV (200 nm) (see Fig.~\ref{fig:opticsboth}), which implies that MnF\textsubscript{2} should
%be optically transparent and should not have any specific color. However, MnF\textsubscript{2} is known to have faded pink color.

\subsection*{Spin-flip vertex as the fundamental component for the excitons in the visible range in MnF\textsubscript{2}:}

The 6.2\,eV excitation is dark in AFM QS$G\widehat{W}$ and bright in paramagnetic QS$G\widehat{W}$, similar to the
1.6\,eV excitation in NiO. However, to explain the pink color of MnF\textsubscript{2} some
absorption in the visible range is necessary, which is not captured by QS$G\widehat{W}$.  It is, however, captured by
DMFT: using non-magnetic QS\emph{GW}+paramagnetic DMFT we find a series of excitonic peaks in Im$\chi_{c}$ that purely
originate from the dynamic vertex in DMFT (Fig.~\ref{fig:opticsboth}).  This is naturally understood from the atomic
configuration of MnF\textsubscript{2}; a local charge excitation is allowed only when a spin-flip component
$\Gamma_{\uparrow\downarrow}$ is present in the charge vertex (see Fig.~\ref{fig:green}). This is why the charge
excitations in the visible range occur only because a \emph{high-order, dynamical vertex} can preserve the atomic spin
configuration on average but at any moment in time it allows for transitions between different spin states (from
$\left<S\right>$=5/2 to $\left<S\right>$=3/2,1/2).  With these series of optical absorption we observe that MnF\textsubscript{2} appears light-red in
color (see Fig.~\ref{fig:green}). The theoretical peak energies agree remarkably well with the experimental optical
absorption spectrum~\cite{suganobook,stout1959} in the visible range. The color 
generated from $\chi_c$ by DMFT is very close to pink (the exact color is very sensitive to the
precise details of the intensities and positions of these peaks). Our observations are completely consistent with a
prior two-band Hubbard model calculation~\cite{color2010}.  There it was shown that for two
half-filled Hubbard bands, a charge peak appears at $\sim{2}J$.  Within such a model the only allowed inter-orbital
transition involves a spin-flip mechanism.  Such optical transitions can occur via a spin-flip vertex, although such a
$d{\rightarrow}d$ transition would be forbidden in an atom.  This two-band model is free of any other candidate symmetry
breaking mechanisms usually considered that relaxes the Laporte rule, and it seems to contain the essential principles
in our detailed \emph{ab initio} calculation, which well reproduces the observed absorption spectrum.
The intensity of the observed peaks within DMFT are arbitrarily multiplied by a large factor to bring
  them to the same scale as experimental peaks. Since the symmetry lowering mechanisms are absent from our approach, we
  can not account for the brightness of these peaks. To put it in perspective, the intensities of these peaks from DMFT
  approach, in both NiO and MnF\textsubscript{2}, are at least 6-7 orders of magnitude weaker than the peaks in transition-metal
  dichalcogenides and 2-3 orders weaker than the peaks in CrI\textsubscript{3}. However, as we have explained in the previous
  sections, there is no reason why intensities of our computed excitonic peaks in optical absorption should agree with
  the intensities of the peaks observed in IXS or optical emission spectroscopy.

%\section*{Conclusions}

Strongly correlated electronic systems with a large fundamental gap can exhibit colors only because of collective charge
excitations deep within the gap, which information is contained in the vertex in the two-particle
  susceptibility.  Understanding origins of these collective excitations in a specific material can pose a formidable
challenge.  For transition metal compounds such as NiO and MnF\textsubscript{2} studied here, these excitations
originate from dipole-forbidden $d{\rightarrow}d$ transitions.  By applying two complementary high fidelity, \emph{ab
  initio} approaches to the vertex function, we are able to fully explain the observed subgap absorption spectra and the
colors correlative to them.  Both approaches, a perturbative approach with a non-local but low-order
vertex, and a local vertex with all graphs included, can produce the desired collective charge
excitations in NiO that determines its green color. However, the deep excitons in
    MnF\textsubscript{2} responsible for its pink color are not captured by low-order perturbation theory.  The locally
    exact method succeeds because it is a nonperturbative theory, which contains high-order spin flip diagrams that also
    contribute to the charge vertex.  Spin disorder has only a slight effect on the one-particle Green's function, and
    similarly on the exciton energy levels that emerge from the two-particle Green's function.  We showed, however, spin
    disorder serves as a symmetry-breaking mechanism, one among several such as spin-orbit coupling, vibronic
    mechanisms, $pd$ hybridization, and defects, all of which affect the exciton's brightness by enhancing their
    oscillator strengths.

 Since the normally large dipole matrix element is absent by symmetry, the relative strength of such
   symmetry-breaking mechanisms that give rise to smaller matrix elements are important in the determination of the
   exact color, since they control intensities of the peaks in the visible range.  However, the
presence and energetics of these excitonic states are largely
 independent of those mechanisms, including spin disorder, and only sensitive to the adequate physical
 component to the vertex function, screened coulomb exchange and its spin components, which we have in our theory. We
 believe, this is a step change in the study of collective charge excitations of strongly correlated systems in that we
 are able to disentangle key mechanisms responsible for the emergence of the excitons and mechanisms that control their
 brightness. In the process, we establish the viability and limitations of two of the most effective approaches in studying collective excitations \emph{ab initio}; DMFT and GW+BSE.

Study of collective charge excitations has come to the forefront of interesting physical phenomena in
  diverse fields in recent years, with advances in resonant, non-resonant, elastic and in-elastic X-ray scattering
  measurements that are combined with atomic force microscopy and scanning tunneling microscopy. Further, understanding
  atomic-like excitons and their spin components have been at the forefront of the research in
  optoelectronics~\cite{novoselovcri3,opto}, single-photon emitters and even qubits~\cite{qubit,qubit1,qubit2}. We
  believe, our work provides a timely guide to this field. The fact that we can compute these excitonic eigenvalues and
  eigenfunctions without any subscription to the atomic ligand field theory, we believe, is a desired step in the right
  direction. What we believe we understand can change or lead to enhanced understanding
  only when we are able to compute observables explicitly. We show, being able to compute these excitons and their spin
  components, provide us with that key understanding of these systems and the desired theoretical methods for them, in
  parallel to solving an long-standing problem of essentially many-body nature.

\section*{Methods}

\subsection*{Simulations of the ordered anti-ferromagnetic phase: LDA, QS\emph{GW} and QS$G\widehat{W}$ self-consistency:}

For the simulations of the ordered AFM phase 4 and 6 atom unit cells of NiO and MnF\textsubscript{2} were used
respectively. Single particle calculations (LDA, and energy band calculations with the static quasiparticlized
QS\emph{GW} self-energy $\Sigma^{0}(k)$) were performed for both NiO and MnF\textsubscript{2} on a 12$\times$12$\times$12
\emph{k}-mesh  while the (relatively smooth) dynamical
self-energy $\Sigma(k)$ was constructed using a 4$\times$4$\times$4 \emph{k}-mesh and $\Sigma^{0}$(k) extracted from it.
For each iteration in the QS\emph{GW} self-consistency cycle, the charge density was made self-consistent.  The
QS\emph{GW} cycle was iterated until the RMS change in $\Sigma^{0}$ reached 10$^{-5}$\,Ry.  Thus the calculation was
self-consistent in both $\Sigma^{0}(k)$ and the density.  Numerous checks were made to verify that the self-consistent
$\Sigma^{0}(k)$ was independent of starting point, for both QS$GW$ and QS$G\widehat{W}$ calculations; e.g. using LDA or
Hartree-Fock self-energy as the initial self energy for QS\emph{GW} and using LDA or QS\emph{GW} as the initial
self-energy for QS$G\widehat{W}$. In the two-particle hamiltonian 16 valence and 16 conduction bands were incorporated
to achieve convergence. The dielectric response is converged on a k-mesh of 4$\times$4$\times$4.

\subsection*{Simulations of the paramagnetic phase: QS\emph{GW} and QS$G\widehat{W}$ self-consistency:}

For the disordered paramagnetic phase the supercell size is 32 for NiO and 48 for MnF\textsubscript{2}.  The supercell
paramagnetism is simulated using special quasi-random structure prescription of Alex Zunger~\cite{sqs}. Nearest
neighbor, two-body nearest neighbor and three body nearest neighbor correlation functions are made in a way to match
exactly that of a random alloy. For NiO and MnF\textsubscript{2} QS\emph{GW} and QS$G\widehat{W}$ calculations were performed on a
2$\times$2$\times$2 k-mesh for self energy. Solution of the two-particle Hamiltonian is converged with 96 valence and 96
conduction bands.

\subsection*{Paramagnetic DMFT calculations using ED and CTQMC impurity solvers:}

For DMFT both ED and CTQMC solvers are used.  CT-QMC calculations are performed on five \emph{d}-orbitals of the
3\emph{d} transitions metals. For NiO, while performing nonmagnetic QS\emph{GW}+paramagnetic DMFT, the Hubbard model in
DMFT is solved using U=8\,eV, $J$=1.0\,eV and double counting correction of 55\,eV as proposed in a previous
work~\cite{sasha-nio}. This allows us to produce exactly 4\,eV of one-particle gap in paramagnetic DMFT. Once the
one-particle experimental gap is reproduced, we explore the local vertex corrected dynamic susceptibilities in charge
and spin channels inside the one-particle gap. For MnF\textsubscript{2}, the Hubbard model on the five Mn 3\emph{d} orbitals is
solved within DMFT using U=16\,eV, $J$=1.3\,eV and double counting correction of 68\,eV. Note that in both the cases the
used double counting correction is slightly smaller than the fully localized limit, for the reasons explained in detail
in this earlier work~\cite{sasha-nio}. When ED is used as the impurity solver, the bath parameters are chosen in a way
that reproduces the fully self-consistent CTQMC+DMFT self energy. For ED, both the Lanczos and exact brute force ED
calculations are performed. The essential collective charge excitations are completely independent of the solver. It is
the locally exact nature of the vertex in DMFT in the paramagnetic phase which is the key for these in-gap charge
excitations.

\section*{Data Availability}

We have created an online repository with all input/output data files that are relevant to run calculations and reproduce all relevant results from the paper. All the input file structures and the command lines to launch calculations are rigorously explained in the tutorials available on the Questaal webpage~\cite{questaal_web} \href{https://www.questaal.org/get/}.

\section*{Code Availability}
The source codes for LDA, QS\emph{GW} and QS$G\widehat{W}$ are available from~\cite{questaal_web}  \href{https://www.questaal.org/get/}  under the terms of the AGPLv3 license.

\section*{Acknowledgments}

MIK, AIL and SA are supported by the ERC Synergy Grant, project 854843 FASTCORR (Ultrafast dynamics of correlated electrons
in solids). MvS and DP were supported the Computational Chemical Sciences program within the Office of Basic Energy
Sciences, U.S. Department of Energy under Contract No. DE-AC36-08GO28308.  This research used resources of the National
Energy Research Scientific Computing Center (NERSC), award
BES-ERCAP0021783, under DOE Contract No. DE-AC02-05CH11231.
 We acknowledge PRACE for awarding us access
to Irene-Rome hosted by TGCC, France and Juwels Booster and Cluster, Germany. This work was also partly carried out on
the Dutch national e-infrastructure with the support of SURF Cooperative. 

\section*{Author Contributions}
MIK and AIL conceived the main theme of the work. SA, MvS, DP have carried out
the calculations. CW has contributed the ED codes. All authors have contributed to the writing of the paper and the analysis of the data.

\section*{Competing interests}
The authors declare no competing financial or non-financial interests.
\section*{Correspondence}
All correspondence, code and data requests should be made to SA.

%\bibliographystyle{ieeetr}
%\bibliography{./gw}
%\bibliography{./tise2,./gw,./mvs-paps,es}

\section*{Supplementary Discussion}

The supplemental materials discuss the Questaal implementation, band structural details from different
levels of the theory, the role of higher order vertex corrections, incorporated in a many body perturbative approach, in
leading a systematic improvement in quasi-particle description and collective charge excitations in these strongly
correlated systems. It also discusses the paramagnetic band structures and the convergence of the imaginary part of the
macroscopic dielectric response $\mathrm{Im}\,\epsilon$ with different sizes of the BSE Hamiltonian.

A detailed justification for why QS\emph{GW} improves on conventional forms of 1-shot \emph{GW} is
	described in \S4 of Questaal's methods paper, Ref.~\cite{questaal_paper} (heretofore referred to as Ref.\,I); in 
	the paper describing Questaal's implementation of QS\emph{GW}, Ref.~\cite{Kotani07} (heretofore referred to as
	Ref.\,II) and finally in Ref.~\cite{Cunningham20}, {\S}IIC (heretofore refereed to as Ref.\,III).  The enhancements
	DMFT brings when augmenting QS\emph{GW} is presented in several works; see, e.g. (I, \S5), and
	Refs.~\cite{acharyasro2021,AcharyaFeSenematicity}.

\subsection*{Questaal Implementation of MBPT}

Implementation of \emph{GW} requires both a 1-body framework and a two-body framework.  Both are described in detail I
	and II.  I places heavier focus on the one-body part, while II focuses on the \emph{GW} theory and its implementation.
	Ref.\,III explains how ladder diagrams are implemented in both the polarizability and self-energy in many-body
	perturbation theory, and provides extensive benchmarks of results.

	Questaal is an all-electron method, with an augmented wave basis consisting of partial waves inside augmentation spheres, constructed from numerical solutions
	of the radial Schrodinger equation on a logarithmic mesh (I, \S2.2). The one-body basis consists of a linear combination of smooth, atom-centered Hankel
	functions as envelope functions, augmented by the partial waves.  Two partial waves are calculated at some linearization energy $\phi_\ell$ and energy
	derivative $\dot{\phi}_\ell$, which provides enough freedom to match value and slope to the envelope functions (I, \S3).

	\emph{One particle basis}: In a conventional LMTO basis, envelope functions consists of ordinary Hankel functions,
	parameterized by energy $E$.  Questaal's smooth Hankel functions are composed of a convolution of Gaussian functions of
	smoothing radius $r_{s}$, and ordinary Hankel functions (I, \S3.1); thus two parameters are needed to define the
	envelope.  In the periodic solid, Bloch sums of these functions are taken (I, Appendix C).  In the present work, $E$ is
	constrained to a fixed value ($-0.4$\,Ry), and $r_{s}$ determined by optimizing the total energy of the free-atomic wave
	function.  These are kept fixed throughout the calculation, while the partial waves and linearization energy float as
	the potential evolves.  By fixing $E$ to a universal value, we are able to take advantage of the ``screening
	transformation'' to render the basis set short-ranged (see I, \S2.9).  This can be useful for the interpolation of the
	self-energy to an arbitrary $k$ mesh, as described below.  A second envelope function of a deeper energy is needed to
	make the hamiltonian reasonably complete.  The latter energy is chosen to be 0.8\,Ry deeper than the first.  The
	envelopes of orbitals $l{=}0{\dots}4$ are employed the first energy (25 orbitals), and $l{=}0{\dots}3$ for the second.
	At the \emph{GW} level, a few other additions are made to make the basis more complete.  
	%Completeness of the envelope functions is sometimes improved by adding ``floating orbitals'' --- points in the
	%interstitial regions where smooth Hankel functions are placed without an augmentation sphere (I, \S3.11), usually for
	%$\ell$ up to 2.  $N_{\rm flt}$ in the Table indicates how many points in the unit cell where floating orbitals are
	%added.
	To expand the hilbert space inside the augmentation spheres, a local orbital $\phi_z$ may be added (I, \S3.7.3).
	$\phi_z$ is a solution of the radial Schrodinger equation at an energy, either well below the linearization energy for
	deep core-like states, or well above it to better represent the unoccupied states.  In the present work we used
	the Ni and Mn 4{d} local orbital.

	\emph{k convergence}:
	The $GW$ mesh and the one-body mesh are generally different: the latter normally needs to be somewhat finer, as the
	self-energy is a relatively smooth function of $k$ while the kinetic energy is less so.  Since the cost is low, we use a
	finer mesh than necessary for the one-particle part, which obviates the need to test the mesh for $k$ convergence.
	Careful tests of the $GW$ mesh were made for each system.  For the simple AFM structures we used a $4{\times}4{\times}4$ 
	mesh for the self-energy, and an $12{\times}12{\times}12$ mesh for the one-particle part.
	for the PM structures (volume increased by 8-fold) the meshes were reduced to $2{\times}2{\times}2$ and $4{\times}4{\times}4$.

	To enable inequivalent meshes, the self-energy must be interpolated.  To render the interpolation everywhere smooth, (I,
	\S2G) eigenfunctions and self-energy are rotated to the LDA basis, and the full self-energy matrix is kept only up to a
	cutoff above the Fermi level in this basis, denoted $\Sigma_\mathrm{cut}$ in the Table.  Above this cutoff, only the
	diagonal part of $\Sigma$ is kept.  $\Sigma_\mathrm{cut}$ may be made arbitrarily high, but if it is too high the
	interpolation is no longer smooth.  Fortunately the result depends weakly on $\Sigma_\mathrm{cut}$.  In this work
	we used $\Sigma_\mathrm{cut}{=}2.5$\,Ry.

	A smooth Hankel functions has a plane-wave representation; thus any linear combination of them, e.g., an eigenfunction,
	does also.  An eigenfunction represented in this form is equivalent to a representation in an LAPW basis: it is defined
	by the coefficients to the plane waves, the shape of the partial waves and their coefficients (which are constrained to
	match smoothly onto the envelope functions).  We used 6.2\,Ry as PW cutoff for the one-particle basis.

	\emph{Two-particle basis}: The two-particle basis is needed to represent quantities such as the bare coulomb interaction
	and the polarizability.  As with the one-particle basis, it as a mixed construction with interstitial parts and
	augmentation parts (II, \S{II}A): envelope function products are represented as plane waves, since product of plane
	waves is another plane wave.  Thus the interstitial parts of the mixed (product) basis are plane waves, with its own PW
	cutoff.  We used 5.0\,Ry as the cutoff.  Inside augmentation spheres, all possible products of
	partial waves are called product functions $B_\ell$, organized by $\ell$ with a form
	$B_I{=}B_{\ell}(r)Y_{{\ell}m}(\mathbf{\hat{r}})$.  The set of all possible products of partial waves is somewhat
	overcomplete with a relatively large rank. It is reduced by diagonalizing the overlap matrix, and retaining the subset
	of functions above a cutoff eigenvalue of the overlap.  It has been found from experience that eigenfunctions with
	eigenvalues below $3{\times}10^{-4}$ for $\ell$=0,1 and $10^{-3}$ for $\ell{>}1$ have essentially negligible effect on
	the result, and are discarded.  The product basis is truncated at a finite $\ell$; we used
	$\ell_\mathrm{cut}{=}8$ in this work.

	\emph{GW: Bare coulomb interaction}: To stabilize the calculation, the bare coulomb interaction, $v(q)=1/q^2$, is
	approximated by a Thomas-Fermi form, $v(q)=1/(q^2+V_\mathrm{TF})$.  This is because if $V_\mathrm{TF}$ is set to zero,
	the result can become unstable.  We use a small value $V_\mathrm{TF}$, typically $2{\times}10^{-5}$\,Ry, though
	sometimes somewhat larger values, up to $2{\times}10^{-4}$\,Ry were used.  The dielectric constant,
	$\epsilon_\infty$, can vary by a few percent over this range.  For that reason $\epsilon_\infty$ was calculated for
	several values of $+V_\mathrm{TF}$, e.g. $1{\times}10^{-5}$, $1{\times}10^{-5}$, and $3{\times}10^{-5}$\,Ry, and the
	reported value is the result when extrapolated to zero.

	\emph{Frequency mesh}: to construct the self-energy, an energy integration on the real frequency axis is taken.  A
	regular quadratic mesh of the form $\omega_i = \texttt{dw}{\times}i + \texttt{dw}^2i^2/(2\omega_c)$ is used, with $i$
	spanning $\omega_i$=0 and the largest eigenstate.  Points are linearly spaced for $\texttt{dw}\ll\omega_c$, but the
	spacing increases for $\texttt{dw}\gtrsim\omega_c$.  It has been found empirically that results are essentially
	independent of mesh for $\texttt{dw}{<}0.08$\,Ry and $\omega_c{\gtrsim}0.1$\,Ry.  In practice we use
	$\texttt{dw}{=}0.02$\,Ry and $\omega_c{=}0.2$\,Ry to obviate the need for checking convergence.  To pick up the poles of
	$G$ and $W$ to make $\Sigma$, the contour is deformed to include an integration on the imaginary axis of $\omega$ (I,
	\S2F).  In all the calculations used here, we used 6 points on a Legendre quadrature.  A few checks showed that the
	result hardly depended on the number of points in the quadrature.

 \emph{Manual vs auto-generated input}: Questaal has an automatic generator, \texttt{blm}, to construct
	input files from structural data.  Most input parameters are automatically generated by \texttt{blm}, such as the MT
	radii $r_\mathrm{MT}$, the product basis cutoffs, and the plane wave cutoffs, the Gaussian smoothing radius defining
	the envelope functions, and the placements for floating orbitals, when they are sought.  Also for the vast majority of
	parameters, the code uses default values if inputs are not explicitly specified.  For a few parameters, manual
	intervention is needed to monitor convergence, especially the number of $k$ points and the plane wave cutoffs.  Hankel
	function energies $E$ must be manually set, but usually fixed values noted above are sufficient.  Results are largely
	insensitive to the choice of $E$, provided it is not pushed too deep.

\subsection*{NiO: $p-d$ alignments and band gaps at different levels of the theory}

\begin{figure}
	\begin{center}
		\includegraphics[width=0.3\columnwidth]{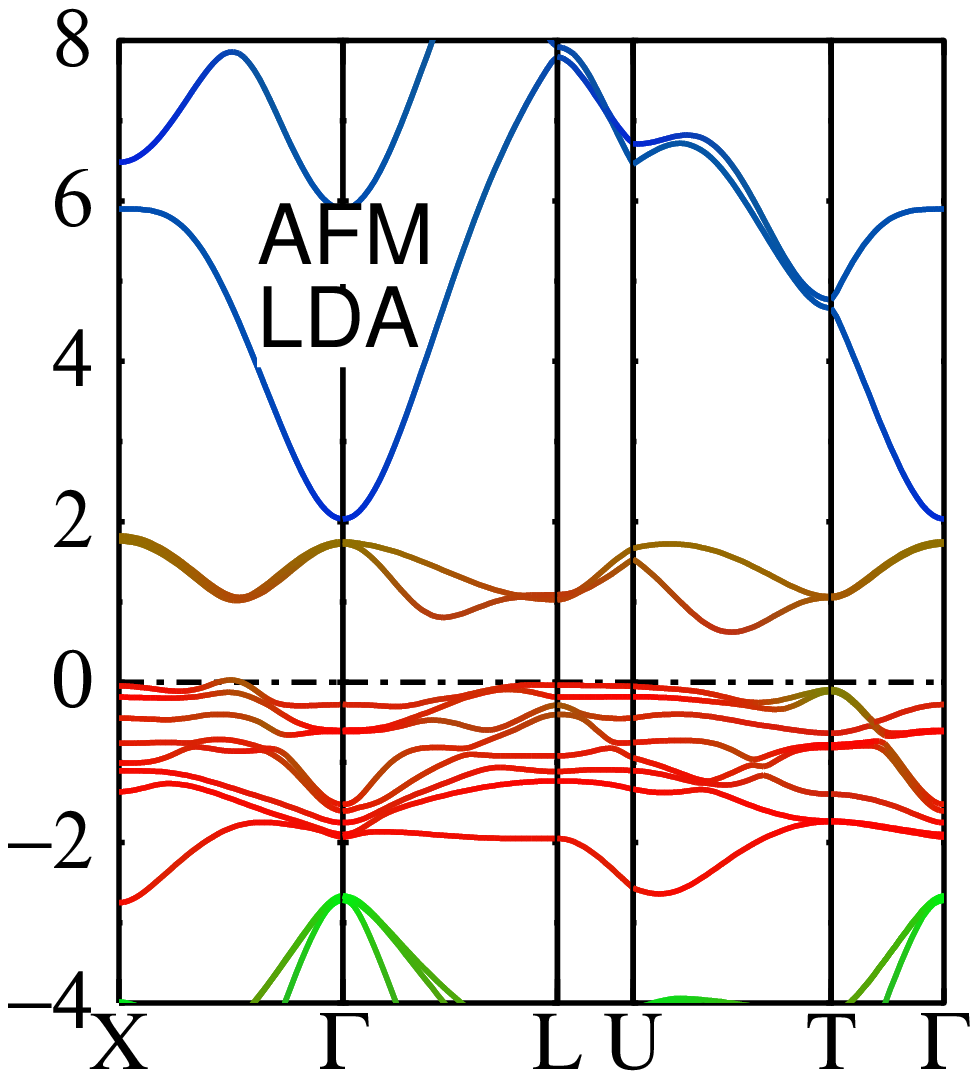}\ 
		\includegraphics[width=0.3\columnwidth]{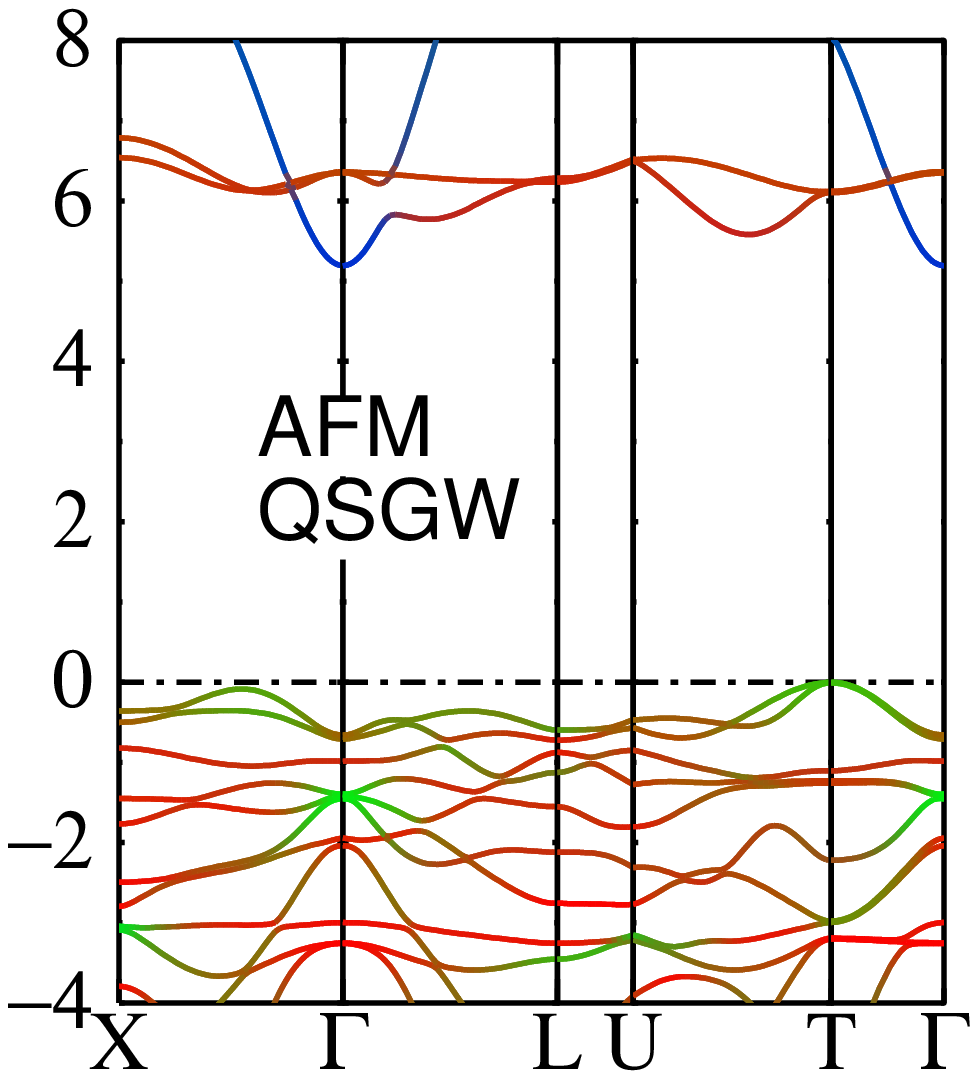}\
		\includegraphics[width=0.3\columnwidth]{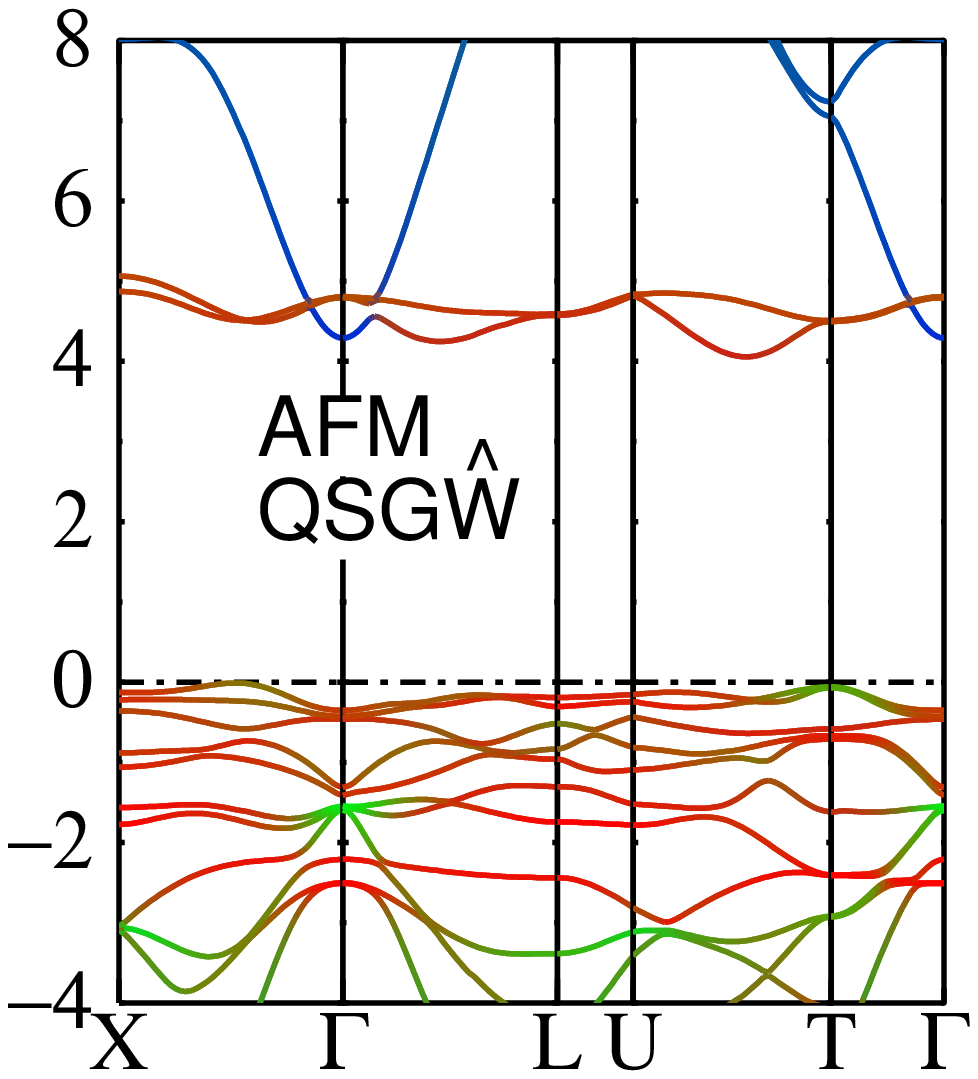}
		\caption{{\bf Electronic band structure of NiO}: Top panels show energy band structures for antiferromagnetic NiO,
			in the local-density, QS$G{W}$, and QS$G\widehat{W}$ approximations.  Blue and red correspond to Ni \emph{sp} and
			\emph{d} character, respectively; green to O character.  The electronic gap is 0.5\,eV in LDA, 5\,eV in
			QS\emph{GW} and 4.0\,eV in QS$G\widehat{W}$.}
		\label{fig:bandnio}
	\end{center}
\end{figure}

Supplementary Fig.~\ref{fig:bandnio} compares three levels of theory for the band structure of NiO.  The self-energy
corrections in QS\emph{GW} significantly modify the $p{-}d$ alignment and also the t$_{2g}{-}$e$_{g}$ alignment relative to
the LDA.  In the LDA the splitting between occupied and unoccupied \emph{d} levels is severely underestimated, and the O
2\emph{p} is severely misaligned with the Ni 3\emph{d} states.  QS\emph{GW} tends to overestimate the $d{-}d$ splitting (screening
is too small, making the potential too close to Hartree Fock), so the electronic band gap within QS\emph{GW} remains
about 20\% too high compared to experimental observations. QS$G\widehat{W}$ shifts Ni-\emph{sp} states by 0.7\,eV and
also the Ni-e$_{g}$, which leads to a reduction of the QS\emph{GW} gap by $\sim$1.0\,eV. It is important to note the BSE
corrects the electronic eigenfunctions from QS\emph{GW} in an orbital dependent manner.

%\begin{figure}
%	\begin{center}
	%		\includegraphics[width=0.8\columnwidth, angle=-0]{sm1.pdf}\\
	%		\caption{{\bf Electronic band structure of NiO}: On the top panels the red, green and blue are respective the orbital projections to the Ni 3\emph{d} t$_{2g}$, e$_{g}$ and O$_{2p}$ states. The electronic gap is 0.45\,eV in LDA, 5\,eV in QS\emph{GW} and 4.0\,eV in QS\emph{GW}+BSE. Both the $d{-}d$ and $p{-}d$ splitting are severely wrong in LDA and they get systematically bettered in QS\emph{GW} and QS$G\widehat{W}$. In the bottom panels we show the band structures again, but now the Ni-sp states are in blue color.}
	%		\label{fig:bandnio}
	%	\end{center}
%\end{figure}

\subsection*{MnF$_{2}$: LDA, QS\emph{GW} and QS$G\widehat{W}$ compared}

\begin{figure}
	\begin{center}
		\includegraphics[width=0.3\columnwidth]{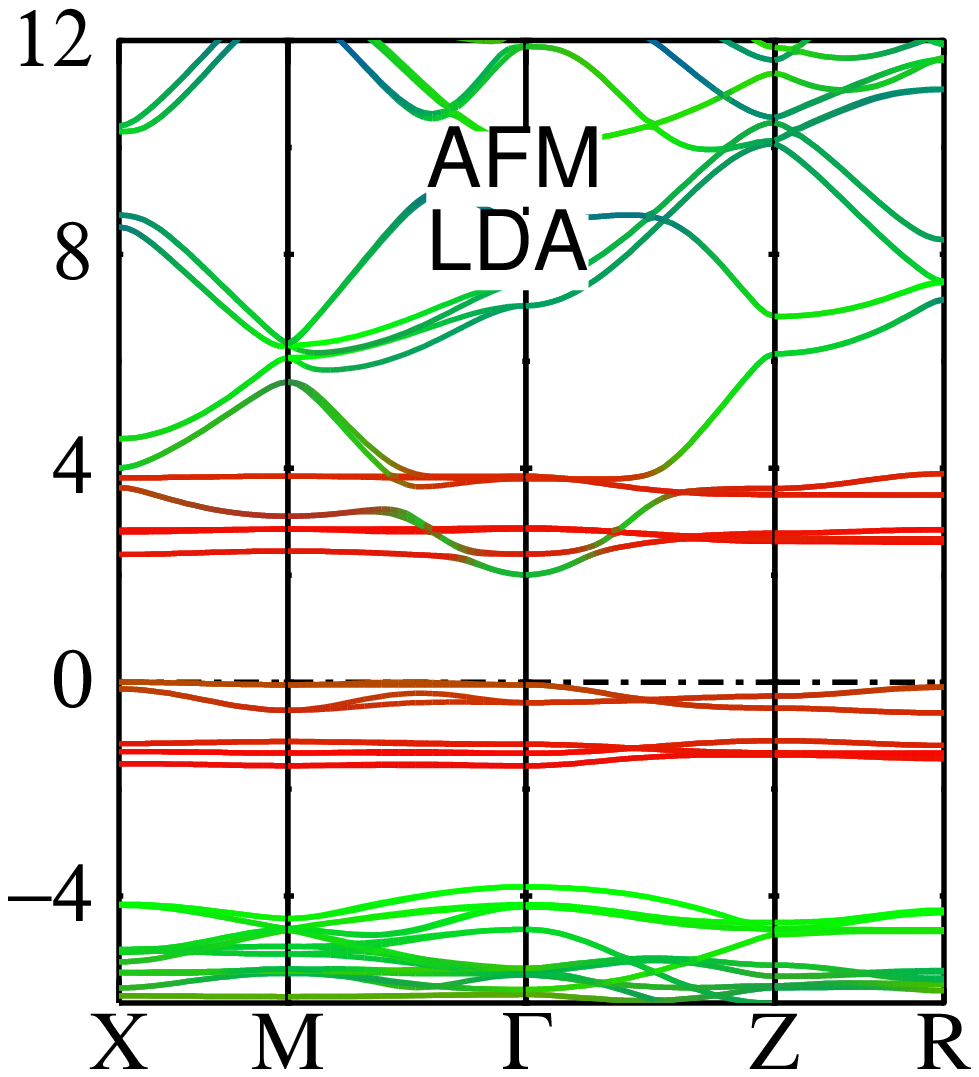}\ 
		\includegraphics[width=0.3\columnwidth]{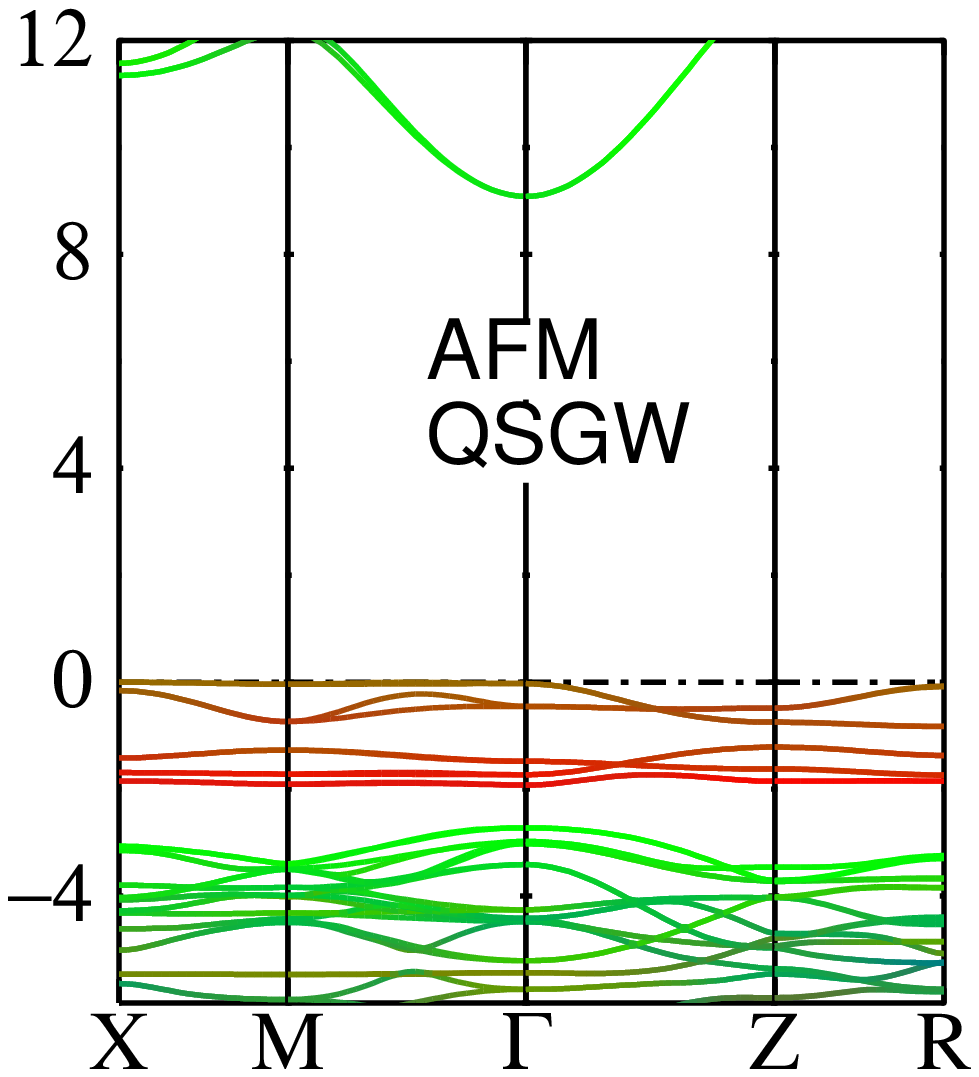}\
		\includegraphics[width=0.3\columnwidth]{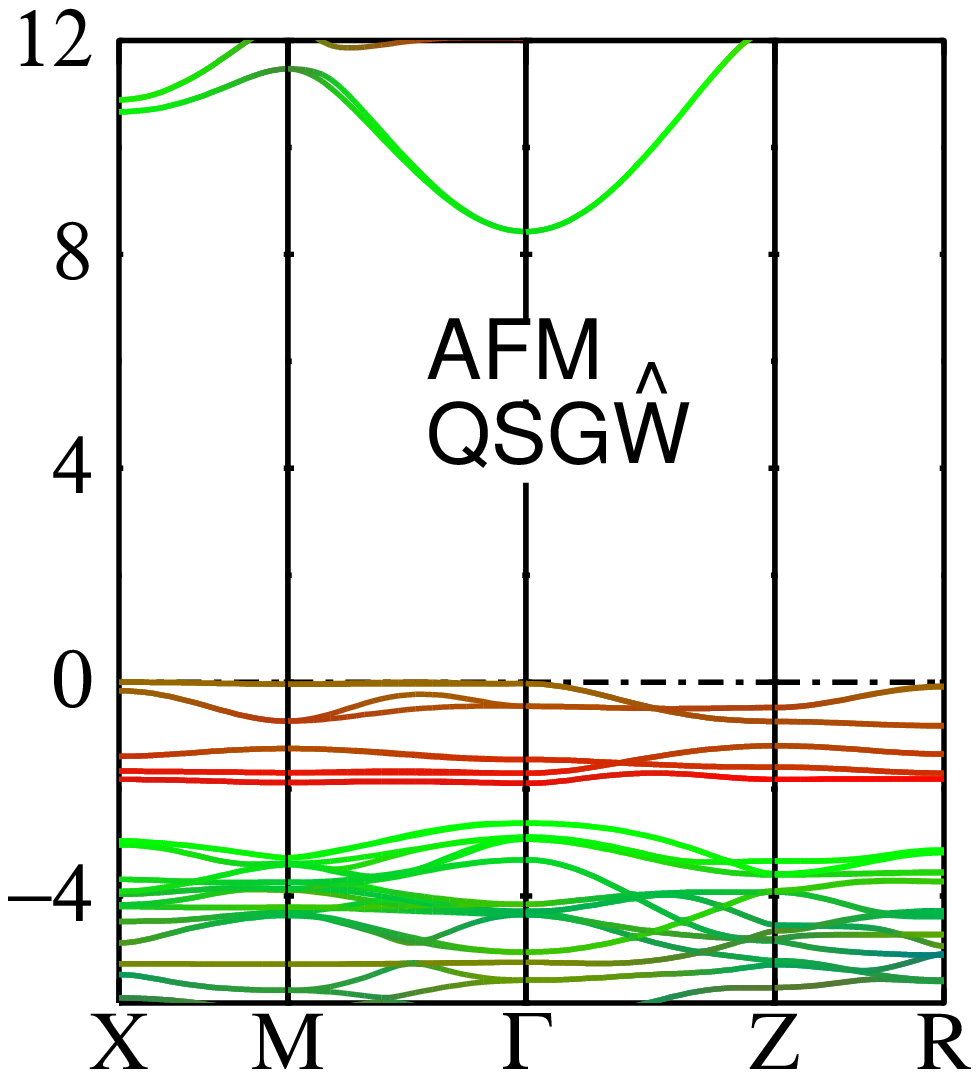}\
		\caption{{\bf Electronic band structure of MnF$_{2}$}, LDA, QS\emph{GW} and QS$G\widehat{W}$.  Red and green correspond to Mn \emph{d}
			(majority spin in the valence band and minority spin in the conduction band), and O \emph{p} character.  The
			electronic gap is 2.0\,eV in LDA, 9.1\,eV in QS\emph{GW} and 8.4\,eV in QS$G\widehat{W}$.  (Had local orbitals been included for the F 3\emph{s}
			and F 3{p} states, the gap would decrease lightly, by about 0.2\,eV.)  QS$G{W}$ (QS$G\widehat{W}$) minority \emph{d} bands lie
			above the figure, at around 14\,eV (12\,eV).}
		\label{fig:bandmnf2}
	\end{center}
\end{figure}

	MnF$_{2}$ provides an extreme instance of errors incurred by the LDA (Supplementary Fig.~\ref{fig:bandmnf2}).  The
	bandgap (2.0\,eV) is very small and the splitting between occupied and unoccupied \emph{d} levels is severely
	underestimated.  Also, the O($2p$)-Mn($3d$) splitting is much greater than in QS$G{W}$. The severe limitations of DFT and also HSE in case of MnF$_{2}$ were discussed in a previous work~\cite{dasmnf2}. QS$G\widehat{W}$ reduces the QS\emph{GW} gap by 0.7 eV.

\subsection*{NiO and MnF$_{2}$: antiferromagnetic and paramagnetic solutions within QS\emph{GW}}

\begin{figure}
	\begin{center}
		\includegraphics[width=0.3\columnwidth]{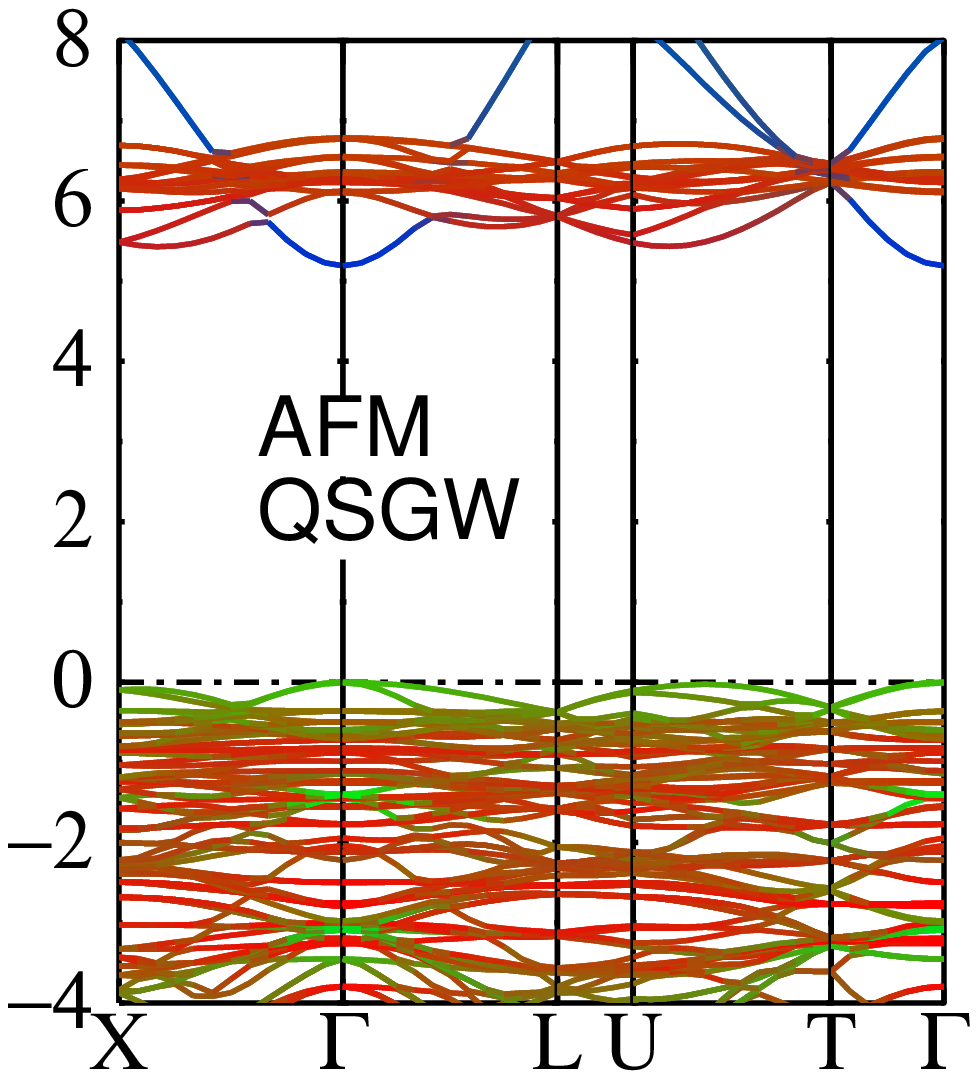}\
		\includegraphics[width=0.3\columnwidth]{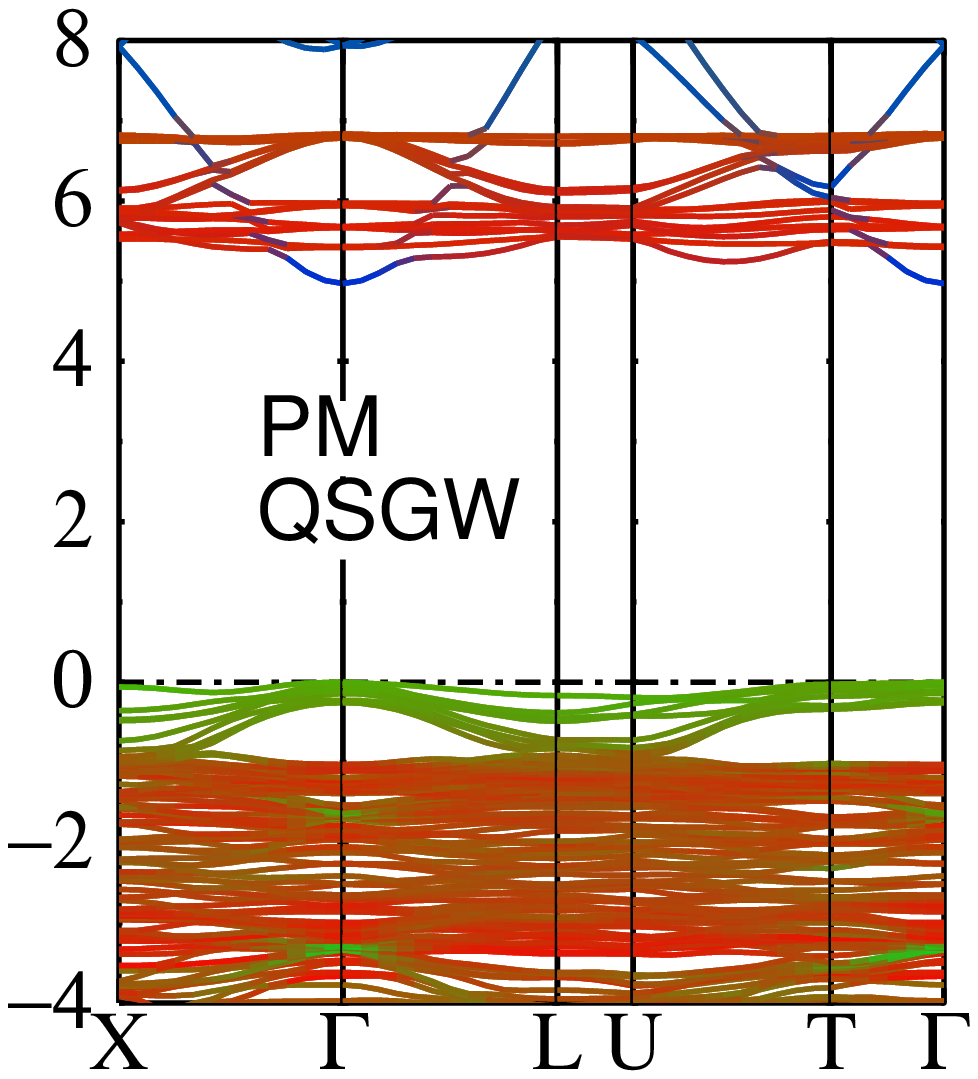}\\
		\includegraphics[width=0.3\columnwidth]{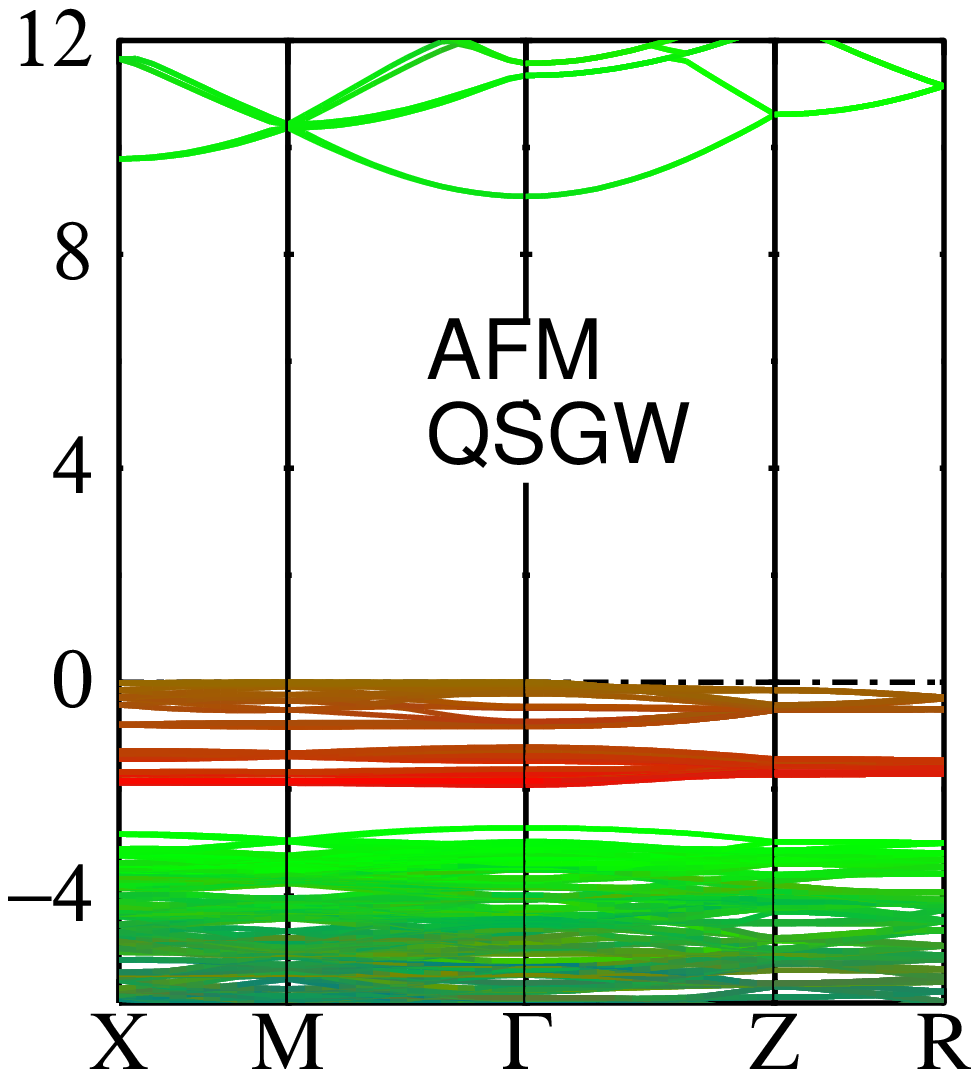}\
		\includegraphics[width=0.3\columnwidth]{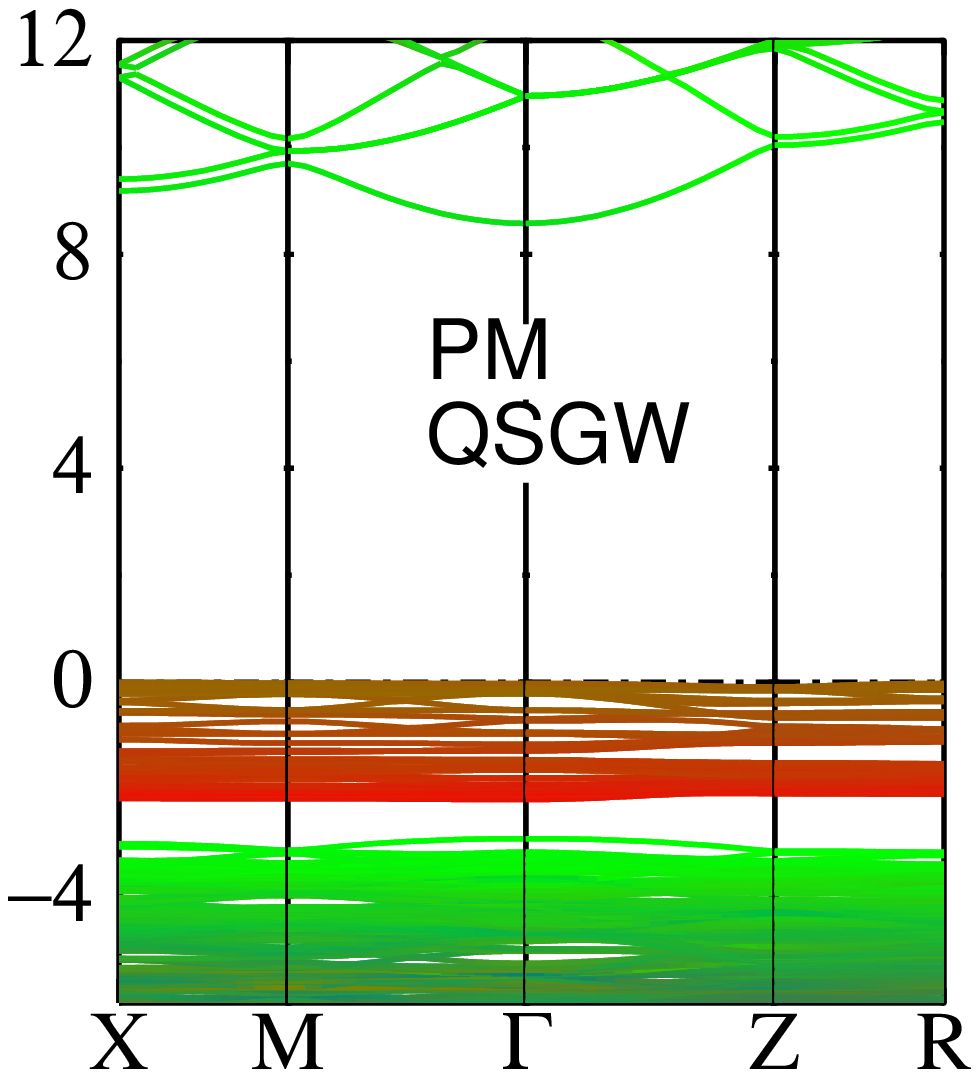}
		\caption{{\bf AFM and paramagnetic phases, Energy bands for NiO and MnF$_{2}$} : Top panels compare
				antiferromagnetic NiO in a 2$\times$2$\times$2 supercell, and paramagnetic NiO in the same cell with quasirandom
				spin arrangements.  Bottom panels are the same for MnF$_{2}$.  The AFM results are equivalent the QS\emph{GW}
				calculations of Figs.~\ref{fig:bandnio} and ~\ref{fig:bandmnf2}, with the Brillouin zone folded.  The red and green colors
				correspond to the projections of one-particle eigenstates onto 3\emph{d} metal (Ni, Mn) and ligand (O,F) respectively.}
		\label{fig:Mott}
	\end{center}
\end{figure}

We simulate the fully ordered antiferromagnetic and disordered paramagnetic phases of NiO and MnF$_{2}$ and compute the
macroscopic dielectric response in both the cases.  A 2$\times$2$\times$2 superlattice of the AFM phase is formed (16 Ni
	or Mn atoms).  To simulate the PM phase, spins are disordered in a quasirandom fashion~\cite{zunger90} so that the
	shortest and second shorted pair correlation functions and three-body correlation functions are equivalent to random
	configurations.  Fig.~\ref{fig:Mott} compares the AFM energy band structure (in the 2$\times$2$\times$2 superlattice) to the
	PM for NiO (top panels) and MnF$_{2}$ (bottom panels).  For NiO, the main change in the AFM$\rightarrow$PM transition is
	a difference in the O($2p$)-Ni($3d$) alignment.  Note how in the PM case the O \emph{p} states are pushed up to
	yield a greater amount of O $2p$ character in the upper valence bands.  There is also a reduction of $\sim$0.2\,eV in the
	bandgap.  The AFM$\rightarrow$PM change is minor in MnF$_{2}$: the gap reduces by $\sim$0.5\,eV.

The AFM$\rightarrow$PM difference is associated with the magnetic ordering energy scales in respective materials. We
observe emergence of several bright excitons in both NiO and MnF$_{2}$ in the paramagnetic phase. However, in NiO, an
otherwise dark exciton in the visible range from the AFM case, becomes bright in the paramagnetic phase while in
MnF$_{2}$ no new excitons emerge in the visible range when the system is magnetically disordered.

\begin{figure}
	\begin{center}
		\includegraphics[width=0.8\columnwidth, angle=-0]{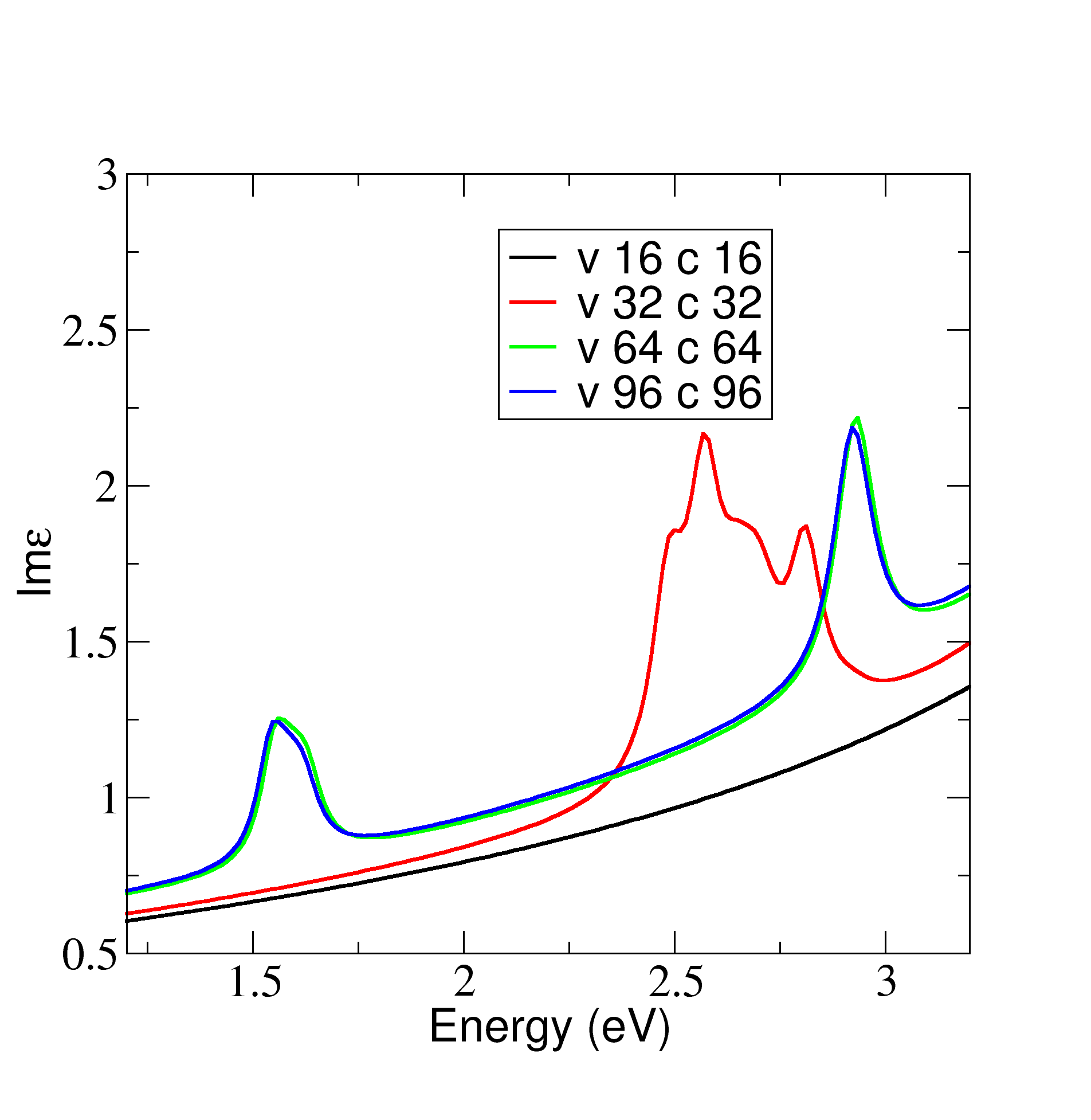}
		\caption{{\bf Convergence in exciton eigenvalues with sizes of the BSE Hamiltonian in paramagnetic NiO} : We show the convergence in the exciton eigenvalues in the paramagnetic QS$G\widehat{W}$ simulation of NiO. We see that inclusion of 64 bands from the valence (v) and conduction (c) allows us to converge two excitonic eigenvalues at 1.6 eV and 2.8 eV simultaneously.}
		\label{fig:nio-afm}
	\end{center}
\end{figure}

Fig.~\ref{fig:nio-afm} shows convergence in excitonic peak positions by solving different sizes of the BSE Hamiltonian.
Beyond 64 valence and 64 conduction bands (included in the BSE Hamiltonian) the eigenvalues stops changing. With
2$\times$2$\times$2 k-mesh and v=64 and c=64, this amounts to solving a BSE eigenvalue problem with matrix of rank
8$\times$64$\times$64=32768.

\end{document}